\documentclass{article}
\usepackage{graphicx} 
\usepackage{xcolor}
\usepackage[utf8]{inputenc}
\usepackage[T1]{fontenc}
\usepackage{hyperref}
\usepackage{tcolorbox}
\usepackage{amsmath}
\usepackage{longtable}
\usepackage{multirow}
\usepackage{booktabs,subcaption,amsfonts,dcolumn}
\usepackage{tikz}
\usepackage{makecell}
\usepackage{float}
\usetikzlibrary{positioning}
\usepackage{color}
\usepackage{listings} 
\usepackage{subfiles} 
\usepackage[style=ieee,dashed=false]{biblatex}
\usepackage[a4paper,width=150mm,top=25mm,bottom=25mm]{geometry}

\definecolor{dkgreen}{rgb}{0,0.6,0}
\definecolor{gray}{rgb}{0.5,0.5,0.5}
\definecolor{mauve}{rgb}{0.58,0,0.82}

\newcommand{\KeY}{\texorpdfstring{Ke\kern-.1emY}{KeY}}

\newcommand{\code}[1]{\lstinline{#1}}

\title{Probabilistic energy profiler for statically typed JVM-based programming languages}
\author{Joel Nyholm$^1$, Wojciech Mostowski$^1$, Christoph Reichenbach$^2$\\
\footnotesize{1. Halmstad University, Sweden}\\
\footnotesize{2. Lund University, Sweden}}
\date{November 2025}

\begin{document}

\maketitle


\begin{center}
\textbf{Abstract}\\
\end{center}

\noindent Energy consumption is a growing concern in several fields, from mobile devices to large data centers.
Developers need detailed information on the energy consumption of their software to mitigate consumption issues.
Previous approaches mainly have a broader focus, such as on specific functions or complete programs, instead of source code statements. Additionally, they primarily focus on estimating the CPU's energy consumption with point estimates, thus disregarding other hardware components' energy consumption and limiting their use for statistical reasoning and explainability. 
We have created a novel methodology to address the limitations of only measuring the CPU's consumption and using point estimates, where we focus on predicting the energy usage of statically typed JVM-based programming languages, such as Java, Kotlin, or Scala.
We measure the energy consumption of Bytecode patterns, the translation from a specific programming language's source code statement to its Java Bytecode representation.
With the energy measurements, we construct a statistical model.
Unlike prior approaches based on frequentist statistics, we create our model via Bayesian statistics, which allows us to predict the expected energy consumption through a statistical distribution and analyze individual contributing factors. 
The model includes three factors we can obtain statically from the code: \emph{data size}, \emph{data type}, \emph{operation}, and one factor about the hardware platform the code executes on: \emph{device}.
To validate our methodology, we implemented it for Java, focusing on Java's Bytecode patterns, and evaluated its energy predictions on previously unseen programs. 
We observe that all four factors are significant, notably that two devices of the same model may significantly differ in energy consumption and that the operations and data types cause significant consumption differences.
The experiments also show that the energy prediction of programs closely follows the program's real energy consumption, further validating our approach. 
Our work presents a methodology (not a tool) to construct an energy model that future work, e.g., energy-aware program analysis or verification tools, can use as their energy estimations.

\section{Introduction}
Energy efficiency is a concern in many areas, such as mobile software design and software in data centers.
Understanding and reducing the energy consumption of even simple software systems is challenging, and modern language run-time systems with dynamic optimizations increase this challenge.

Consider statically typed JVM-based programming languages such as Java, Scala, or Kotlin, which use a platform-independent intermediate representation, Java Bytecode, executed on a virtual machine.
The software's energy consumption depends on the compiler with its optimizations, the Java Virtual Machine (JVM), the garbage collector, the operating system, and the target hardware.
To understand the energy impact of software design decisions, a software engineer would need knowledge that penetrates these abstraction layers and the details of the intended execution platform(s).

Prior work has explored approaches for workload-based energy profiling, where developers run software with a specific workload in a specialized execution environment that measures energy consumption and estimates how much of the consumption can be attributed to individual subroutines or even lines of code~\cite{distributedEnergy,petra,APIEnergy,ANEPROF}.
In practice, workload-based energy profiling requires a custom hardware setup, which limits its suitability for everyday software engineering:
while some hardware platforms expose hardware performance counters for energy use of specific components (e.g., the CPU), few are able to measure the platform's total energy consumption.

\emph{Predictive} approaches for estimating energy consumption avoid this hardware-dependence and rely purely on an execution trace.
For Android, Hao et al.\@ propose such an approach, \emph{eCalc}, which constructs a platform-specific instruction-level energy usage model with five energy cost functions for different operations, and predicts the energy usage of a given execution trace as the sum of the Bytecode instruction estimates~\cite{ecalc}.
Even though their model does not account for interactions across neighbouring instructions (caching, branch prediction, page tables) or more finely-grained programming language constructs, such as data types, Hao et al.\@ report an error of less than 10\% for predicting energy usage of execution traces for several Android benchmarks, albeit on Android 2.x (now obsolete) running on an early Intel Atom CPU (on a custom board made specifically for the study).

Predictive approaches such as \emph{eCalc} are thus appealing as a foundation for software tools that help developers understand the energy usage properties of their code.
However, existing approaches only predict point estimates, such as the \emph{expected value} (the mean over infinitely many runs), which is insufficient for rigorous statistical reasoning or explainability.
These approaches are therefore insufficient for supporting software tools that compare the energy efficiency of two software systems with statistical confidence or estimate the likelihood of whether a system can run for a certain duration on a given battery charge.

To address the limitation of approaches, such as eCalc that only considers the CPU's energy consumption and predicts using point estimates, we propose a new methodology that uses Bayesian statistics to create an energy model that infers a device's total energy consumption when executing statically typed JVM-based programming languages. We here use the term 'methodology' in the sense of a technical procedure that addresses a specific technical challenge. In our case, systematically constructing statistically robust answers to technical questions that can then serve to answer research questions or to construct software tools.
The model our methodology defines uses a quadruple $F=(T,d)$ consisting of a triple $T$, extended by the element $d$. The quadruple explains all factors we model, where the triple explains the source code statement's arguments, $T = \left<o, t, s \right>$ that we can statically obtain from the code: \textbf{o}peration, data \textbf{t}ype, data \textbf{s}ize. The element d is the \textbf{d}evice and can be added if the execution platform is known. It is included for generalization for a specific device, since energy consumption can behave differently on different instances of the same hardware platform due to variations in its electrical components~\cite{pcbVariation,resistorVariation,modernHardwareVariation}. Our model, to our knowledge, is the first predictive energy model for software to explicitly account for the effect of hardware differences on energy consumption. However, the methodology can disregard this category if the software has multiple or no specific target devices.

The model constructs a Gaussian distribution for each quadruple that describes a source code statement, whilst also grouping each quadruple element to describe their individual effects on energy consumption, i.e., the energy consumption effect for operations, data types, data sizes and devices.  
This model approach has several benefits, such as explaining energy effects of changing data types for a given algorithm.

The methodology explicitly targets the worst-case energy consumption (WCEC) of these single code statements and uses the convolution of individual prediction distributions to obtain a prediction for the energy consumption of functions and entire programs. Distributions represent the WCEC because a single point estimate does not explain the possible consumption values that the worst-case scenario can adopt due to random variations in the energy, e.g., because of temperature (Johnson-Nyquist noise) or external interferences (electromagnetic interference). It also retains its statistical properties, such as presenting the mean, skewness, and mode, and enabling calculation of the entropy, cumulative distribution, and probability density functions, which point estimates for WCEC do not retain. 
Using WCEC enables predictable estimation for the worst-case, ensuring that software systems can meet energy specifications, which are vital for battery and low-power devices, and are cost-saving for applications and software executing in data centers. With this information, program analysis or software verification tools can use our methodology as the foundation for their energy estimates.

We have implemented our methodology for a subset of Java, where we focus on Bytecode interpretation, thus executing in interpreter mode. Whilst just-in-time compilation increases performance and lowers energy consumption~\cite{oldJVMarticle}, its compilation during runtime adds unpredictability. As we focus on WCEC, we need a predictable and empirically consistent methodology, which the interpreter mode enables. To validate the suitability of our model for predicting energy consumption, we manually simulate how an abstract interpretation-based program analysis might compute the total energy consumption distribution for several small programs.

The longer-term goal of our research project is to apply static analysis methods
to establish and validate the energy behavior of software systems with the help of declarative annotations, such as those offered by Java Modelling Language~\cite[Chap.~7]{KeYBook2016}.
One possible basis for the static analysis method for energy use is created by Kersten et al.\@~\cite{Kersten2014} and uses a classical programming logic~\cite{Hoare69} approach, and such a technique as proposed in~\cite{Kersten2014} requires an energy consumption model for the programming language and execution platform.

We present our methodology for such a model based on statically typed JVM-based programming languages running on the subject execution platform and measured with suitable equipment for energy consumption.
Our work focuses on one hardware platform, the Raspberry Pi 5, in a controlled setting, but our approach is not specific to any execution platform.

Our work makes the following novel contributions:
\begin{itemize}
    \item A methodology for creating an energy model, providing energy consumption distributions for statically typed JVM-based programming languages.
    \item An investigation of the effect of primitive data types, data sizes, operations, and device instances on energy consumption.
    \item Evidence that different device instances of the same hardware model consume different amounts of energy, affecting the generalizability of energy predictions.
\end{itemize}

The paper's organization is as follows.
Section~\ref{sec:BayesianStatistics} provides background information on Bayesian statistics.
Section~\ref{sec:model} introduces our methodology.
Section~\ref{sec:implementation} presents the implementation of the methodology.
Section~\ref{sec:evaluation} explain the evaluation and validity threats.
The evaluation results are presented in Section~\ref{sec:results}, with the discussion in Section~\ref{sec:discussion}.
Section~\ref{sec:relatedWorks} discusses the related work, and Section~\ref{sec:conclusion} concludes the paper.

\section{Bayesian statistics}
\label{sec:BayesianStatistics}
Bayesian statistics is a statistical framework that aims to make probability statements of a true unobserved phenomenon given prior knowledge~\cite{BDA3,introBayesian,rethinking}.
At a high level, this goal is similar to that of \emph{frequentist} statistics, the statistical framework most common in computer science today.
The main difference between these two frameworks lies in their ways of building and interpreting statistical models: frequentist statistics aims to build models purely from observed data, avoiding biases and assumptions as much as possible.
In contrast, Bayesian statistics builds models that make biases and prior assumptions explicit and susceptible to scrutiny and validation.

We use Bayesian statistics for two reasons: (1) Bayesian statistics forces us to make our assumptions explicit and increase \emph{modularity} and \emph{explainability}, and (2) Bayesian statistics predict energy consumption in the form of \emph{posterior distributions}, which compare energy consumption between different alternatives in more detail compared to single point estimates provided by frequentist approaches.

In this paper, we assume that energy consumption follows a normal distribution and lies within a range that we conservatively select from prior work (Section~\ref{sec:model:prior}). We then pass these \emph{prior distributions} to a \emph{Bayesian inference} algorithm that updates these prior distributions to better explain our measurements.
If we have few measurements, our estimates will dominate the results and may lead to imprecise or even incorrect results.
With enough data, Bayesian inference can correct errors in our priors (outside of pathological cases) and allow us to estimate how well the distribution predicted by our model fits the data (Section~\ref{sec:threats}).

Our methodology uses a \emph{hierarchical model}, which allows combining multiple statistical distributions that separately model different aspects of the underlying statistical process --- allowing us to express statistical independence assumptions.
For example, different instances of the same hardware platform (e.g., different physical CPUs with the same specification) and different instructions will differ in energy consumption, but these variations are independent.

\section{Energy Model Methodology}
\label{sec:model}

\begin{figure}
    \centering
    \includegraphics[width=\linewidth]{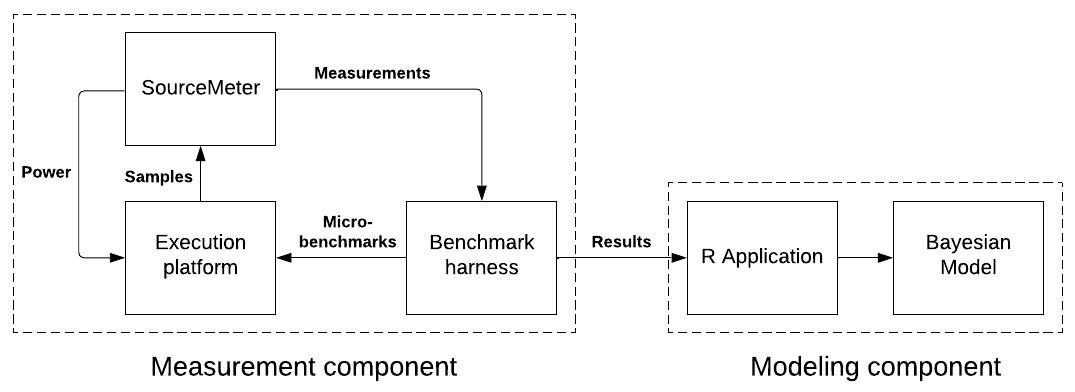}
    \caption{Energy methodology overview}
    \label{fig:EnergyProfiler}
\end{figure}

Our methodology defines a WCEC energy model for a device's total energy consumption, see Fig.~\ref{fig:EnergyProfiler} with two components:
(1)~The measurement component (Section~\ref{sec:model:measurement}), which measures the energy consumption of the benchmark harness executed on the target hardware;
(2)~The modelling component (Section~\ref{sec:model:model}), which creates a Bayesian model of the programming language's energy consumption.

\subsection{Measurement component}\label{sec:model:measurement}
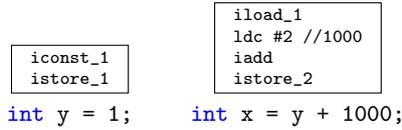
\begin{figure}
  \centering
  \begin{tikzpicture}[N/.style={draw, inner sep=1pt, anchor=south},]
    \scriptsize

    \node[N] (I1) at (0,0) {\begin{tabular}{l}
      \texttt{iconst\_1}\\
      \texttt{istore\_1}\\
    \end{tabular}};

  \node at (0, -0.3) {\small \texttt{\code{int} y = 1;}};

    \node[N] (I2) at (3, 0) {\begin{tabular}{l}
      \texttt{iload\_1}\\
      \texttt{ldc \#2 //1000}\\
      \texttt{iadd}\\
      \texttt{istore\_2}\\
    \end{tabular}};

  \node at (3, -0.3) {\small \texttt{\code{int} x = y + 1000;}};

  \end{tikzpicture}
    \caption{Bytecode patterns for an assignment and one addition, for Java}
    \label{fig:BytecodePattern}
\end{figure}

There are several statically typed JVM-based programming languages.
To enable modelling of these with one approach, we measure the energy of Bytecode patterns, the translation of the programming language's source code statement to its Bytecode representation, as translated by the compiler (see Fig.~\ref{fig:BytecodePattern}).

\subsubsection{Benchmarking setup}
The benchmark setup consists of an execution platform and a multimeter that measures and provides the Voltage $V$ and Amperage $I$ to the execution platform. Since the multimeter measures the electrical properties and the benchmark harness measures the execution time $t$, we can calculate the execution platform's total energy consumption in Joules $J = V \times I \times t$. 

\subsubsection{Benchmark harness}
\label{sec:Method:Meas:BenchmarkHarness}
The benchmark harness contains microbenchmarks, each containing a Bytecode pattern that is classified as a triple $T = \left<o, t, s \right>$ for the operation, data type, and data size. 

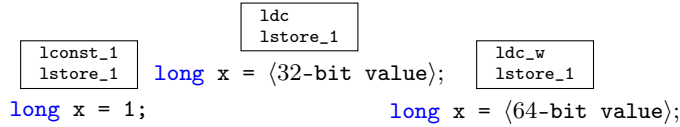
\begin{figure}[H]
  \centering
  \begin{tikzpicture}[N/.style={draw, inner sep=1pt, anchor=south},]
    \scriptsize

    \node[N] (I1) at (0,0) {\begin{tabular}{l}
        \texttt{lconst\_1}\\
        \texttt{lstore\_1}\\
    \end{tabular}};
  \node at (0, -0.3) {\small \texttt{\code{long} x = 1;}};

    \node[N] (I2) at (2.9, 0.5) {\begin{tabular}{l}
        \texttt{ldc}\\
        \texttt{lstore\_1}\\
    \end{tabular}};
  \node at (2.9, 0.2) {\small \texttt{\code{long} x = $\left<32\text{-bit value}\right>$};};
    
    \node[N] (I3) at (6, 0) {\begin{tabular}{l}
        \texttt{ldc\_w}\\
        \texttt{lstore\_1}\\
    \end{tabular}};
  \node at (6, -0.3) {\small \texttt{\code{long} x = $\left<64\text{-bit value}\right>$};};

  \end{tikzpicture}
    \caption{All microbenchmarks for the variable declarations operation, with data type \code{long} and all three possible data sizes for Java}
    \label{fig:microbenchmarkExample}
\end{figure}

A single microbenchmark contains one permutation for a source code statement from the set of all possible data types and data sizes for a given operation (see Fig.~\ref{fig:microbenchmarkExample}). 

Each microbenchmark executes its Bytecode pattern in a loop with $i$ iterations, that ensures that all multimeter measurements happen during the microbenchmark's execution. As the loop and execution platform affect the measurements, a baseline must also be measured. Hence we measure the energy consumption of an empty loop with $i$ iterations, and subtract the mean of the baseline's energy consumption from all measurements to mitigate data noise biases. 

Each microbenchmark measurement happens in the measurement cycle that measures the current, voltage, and execution time for one microbenchmark.
Every measurement cycle begins with configuring the multimeter to prepare it for measuring.
Afterward, to mitigate possible memory related noise (more in Sect.~\ref{sec:threats}), the garbage collector executes to eliminate de-allocations during the measurement cycle.

Once the initialization is complete, the benchmark harness signals the multimeter to measure current and voltage, records the execution's starting time and executes the microbenchmark.
During the execution, the multimeter measures $n_{\textit{meas}}$ current and voltage values.
When the execution concludes, the benchmark harness records the stopping time to calculate the execution time of the microbenchmark.
The execution time is divided by $i$ iterations to obtain the mean execution time of a single microbenchmark execution.
After completing all measurements for the current measurement cycle, the multimeter sends the measurements to the benchmark harness, concludes the ongoing measurement cycle, and waits for the next cycle.

The measurement cycle for each microbenchmark executes 10 times, to account for statistical variation in the measurements, and, use a split plot approach, i.e., randomizing the execution order of the microbenchmark cycles to avoid bias due to ordering.

Since the energy consumption of executing software depends on the hardware, the methodology also focuses on another optional factor, if a specific target device exists: the energy consumption differences between instances of the same hardware platform.
Therefore, the method explicitly includes measurements using $n_{\textit{device}}$ ($>1$) functionally identical instances of the execution platform to increase the model's generalizability.
Hence, the full benchmark harness must be measured $n_{\textit{device}}$ times, once for each hardware instance, where each microbenchmark will record  $n_{\textit{meas}}$  samples 10 times on $n$ execution platforms, resulting in $10\times n_{\textit{meas}} \times n_{\textit{device}}$ energy measurements per microbenchmark.

By including the device category, we represent all model parameters as the quadruple $F=(T,d)$, where $T$ is statically obtainable from the code and $d$ is the device's energy contribution.

\subsection{Modeling component} 
\label{sec:model:model}
The modeling component constructs a Bayesian model from the measurements (Section~\ref{sec:model:measurement}) using the rstan~\cite{rstan} library.

We have no a priori knowledge about the factors that we measure, except that the measurements are continuous and have finite mean and variance.
Therefore, we model energy consumption as a Gaussian distribution (see Eq. \ref{eq:modelFormula_a}), since this distribution does not add other assumptions and is a convenient choice for further modelling operations because it reduces the need for more complex calculations compared to other distributions, speeding up the analysis of our model.
Furthermore, we use a multi-level model (Section~\ref{sec:BayesianStatistics}) with a non-centered parameterization, an optimization that increases our model's efficiency and convergence rate without hindering the result's validity.

\begin{align}
J &\sim \mathit{Normal}(\mu, \sigma) && \text{Likelihood} \label{eq:modelFormula_a}\\
\mu &= \alpha_{datasize[i]} + \beta_{opType[i]} && \text{Likelihood mean} \label{eq:modelFormula_b}\\ 
& \phantom{=} {} + \gamma_{datatype[i]} + \delta_{device[i]} && \nonumber\\
\alpha_j &\sim \mathit{Normal}(\Bar{\alpha}, \sigma_{\alpha}) && \text{Prior for } \mu \label{eq:modelFormula_c}\\
\beta_j &\sim \mathit{Normal}(\Bar{\beta}, \sigma_{\beta}) && \text{Prior for } \mu \label{eq:modelFormula_d}\\
\gamma_j &\sim \mathit{Normal}(\Bar{\gamma}, \sigma_{\gamma}) && \text{Prior for } \mu \label{eq:modelFormula_e}\\
\delta_j &\sim \mathit{Normal}(\Bar{\delta}, \sigma_{\delta}) && \text{Prior for } \mu \label{eq:modelFormula_f}\\
\sigma &\sim \mathit{Exponential}(10^3) && \text{Prior std. dev } \label{eq:modelFormula_g}\\ 
\Bar{\alpha}, \Bar{\beta}, \Bar{\gamma}, \Bar{\delta} &\sim \mathit{Normal}(0.006, 0.001) && \text{Hyperprior} \label{eq:modelFormula_h}\\
\sigma_{\alpha},\sigma_{\beta},\sigma_{\gamma},\sigma_{\delta} &\sim \mathit{Exponential}(10^3) && \text{Hyperprior} \label{eq:modelFormula_i}
\end{align}

\subsubsection{Model Parameters}
We model the energy consumption of one Bytecode pattern as the linear combination of four categories (see Eq.~\ref{eq:modelFormula_b}), from the quadruple $F$ (Section~\ref{sec:Method:Meas:BenchmarkHarness}), resulting in four model categories: operation, data size, data type and device. Each model category's contribution to the final energy prediction for a given Bytecode pattern differs. Accordingly, we assign different distributions to each model category (see Eq.~\ref{eq:modelFormula_c},~\ref{eq:modelFormula_d},~\ref{eq:modelFormula_e}~\&~\ref{eq:modelFormula_f}).

The \emph{operation} category considers the operation of the source code statement, e.g., an array allocation or multiplication.
For each operation, we consider the programming language's different possible \emph{data types}, and \emph{data sizes} for its operand value (e.g., a 32-bit value as the initializer in variable declaration operations).
If the data size parameter is another variable instead of a value, we classify it as a \emph{Load}.
A value corresponding to a constant in Bytecode (e.g. \texttt{iconst\_1}) we classify as a \emph{Constant}.
For loading other values, we distinguish depending on their data size, e.g., \emph{32-bit} or \emph{64-bit} values. The final part of the model is the \emph{device instance}, denoting the $n_i, i>1$ instances the benchmark executes on.
The model includes these categories to create a probability statement of the programming language's energy consumption.

\subsubsection{Prior Distribution}
\label{sec:model:prior}
To make the probability statement, we must define the prior distribution, which is an initial estimate of the energy a given Bytecode pattern consumes. The estimation should stem from prior knowledge, since inaccurate priors will result in misleading inferences~\cite{rethinking}.
We have chosen to model the priors for $\mu$ using hyperpriors (see Eq.~\ref{eq:modelFormula_h}~\&~\ref{eq:modelFormula_i}), whose mean is a Gaussian distribution and whose standard deviation is an exponential distribution.
We model the mean as a Gaussian because it only assumes continuous data with a finite mean and variance. For the standard deviation, we know that it must be a positive value. Since we have little knowledge of the true standard deviations, we use an exponential distribution, a weakly informative prior (an estimate containing a broad assumption about the true standard deviation), reducing the risk of introducing subjective biases in the model.

\begin{figure}[htb!]
    \centering
    \includegraphics[width=0.5\linewidth]{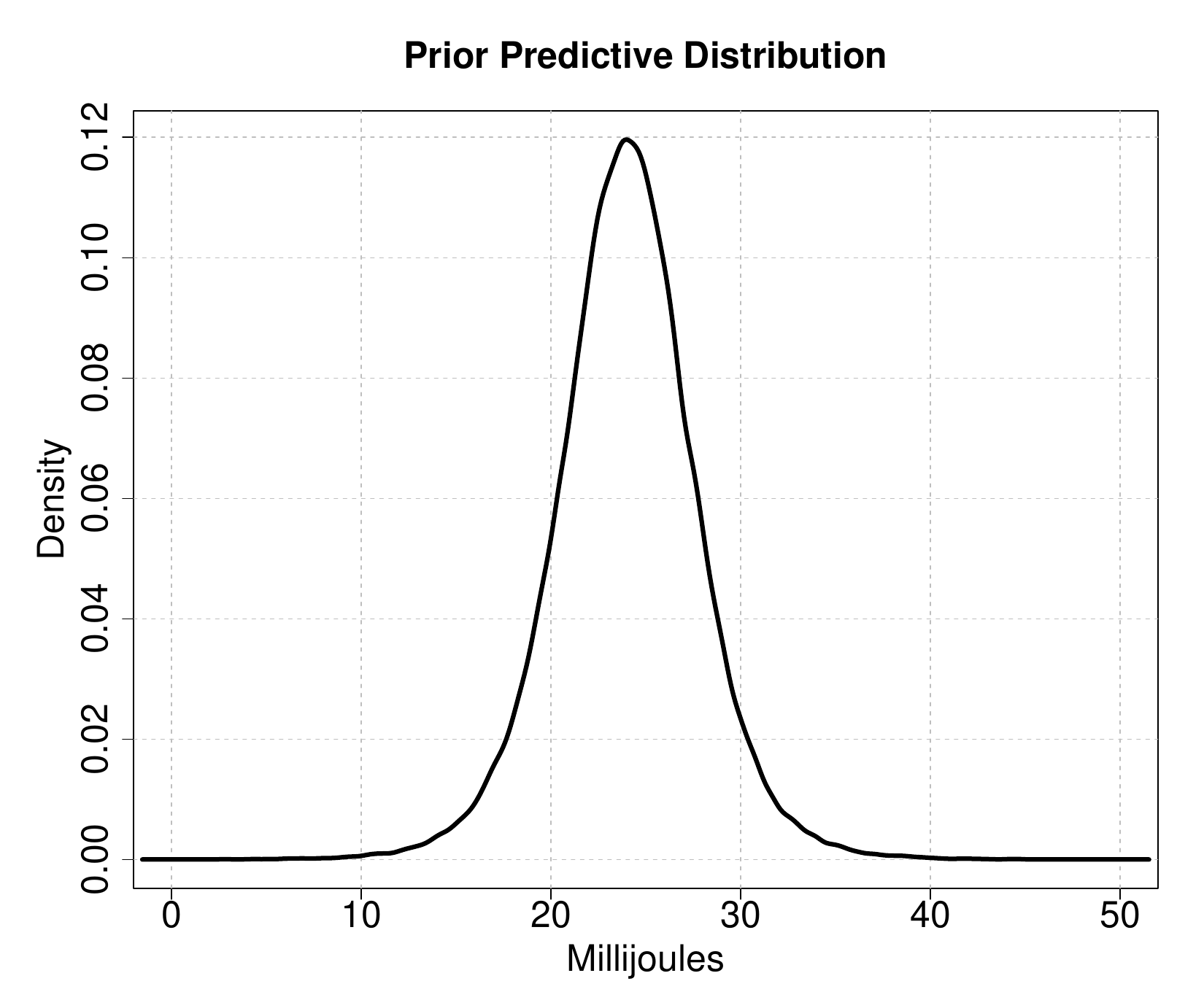}
    \caption{Prior Predictive Distribution of the Energy Model}
    \label{fig:prior}
\end{figure}

Our prior distribution is a Gaussian ranging between 0 and 50 millijoules. We chose the range since related research that measures functions and programs~\cite{ANEPROF,distributedEnergy,vlensApplications,ecalc,petra,APIEnergy}, for Java, presented that smaller functions they investigated consumed between 0 and 260 mJ. Since previous research uses other equipment and measured longer code sequences (functions and programs), we reduced the maximum value from 260 to 50 mJ, an over-approximation of what we believed a single instruction execution would consume. We determined the hyperprior parameters by conducting a prior predictive check (see Fig.~\ref{fig:prior}), simulating which combination of parameters in our model generated the chosen prior distribution of 0--50 mJ.

\section{Implementation}
\label{sec:implementation}
We implement our methodology in Java 21, but the methodology is viable for any statically typed JVM-based programming language.
We focus on a subset of Java (see Tab.~\ref{tab:ModeledBytecodePatterns}) to assess our methodology's viability. All Java source code statements are translated to Java Bytecode by OpenJDK~21.0.6's \texttt{javac} compiler.

\begin{table}
\begin{center}
\resizebox{\linewidth}{!}{%
\begin{tabular}{p{4cm}|lp{1.7cm}p{0.8cm}} 
\toprule
\textbf{Category} & \textbf{Operations} & \textbf{\# Bytecode patterns} & \textbf{Data type(s)}\\
\midrule
\multirow{8}{1.5cm}{\textbf{Arithmetic}} & Addition & 12 & i,l,d,f\\ 
    & Bit Operations & 14 & i,l\\ 
    & Division & 12 & i,l,d,f \\ 
    & Increase & 1 & i\\ 
    & Multiplication & 12 & i,l,d,f\\ 
    & Negation & 4 & i,l,d,f \\
    & Subtraction & 12 & i,l,d,f\\
    & Modulo & 4 & i,l,d,f\\
\hline
\multirow{2}{4cm}{\textbf{Control transfer}} & If statements & 35 & i,l,d,f,r\\ 
    & Switch case & 2 & i\\
\hline
\textbf{Method invocation} & Method call \& return & 6 & i,l,d,f\\ 
\hline
\textbf{Load/Store} & Variable declaration & 12 & i,l,d,f \\
\hline
\multirow{1}{4cm}{\textbf{Object Creation and manipulation}} & Object allocation and field access operations & 17 & i,l,d,f,r\\ 
    & Array allocation and array operations & 16 & i,l,d,f,r\\
\hline
\textbf{Type conversion} & Primitive type conversions & 15 & i,l,d,f\\
\bottomrule 
\end{tabular} %
}
\end{center} 
\caption{Modeled Bytecode patterns, from JVM specification~\cite[Chap.~2.11]{JVMSpec}. i = \code{int}, l = \code{long}, d = \code{double}, f = \code{float}, r = (array or object) reference.
}
    \label{tab:ModeledBytecodePatterns}
\end{table}

\subsection{Benchmarking setup}
Our benchmark hardware setup consists of the execution platform, two Raspberry Pi~5 Model~B revision~1 boards, and the multimeter Keithley 2602 SourceMeter~\cite{Keithley2602Docs} that measures and provides power ($5.6V$ in constant voltage mode) to the execution platform.
In our configuration, the SourceMeter has a voltage resolution of $50\mu V$, an accuracy of $0.02\% + 1.8mV$, and a typical noise of $100 \mu V$. The resolution of the amperage measurement is $10 \mu A$, with an accuracy of $0.06\% + 1.5mA$, and a typical noise of $150 \mu A$, providing us with enough precision and resolution for fine-grained measurements.

The frequency of the Keithley 2602 SourceMeter (50Hz~\cite{Keithley2602Docs}) allows us to record one sample per 20ms.  Additionally, our measurements must account for a latency of 0.12ms per sample we record. Accordingly, having a large sample size per microbenchmark execution becomes infeasible. Our sample size is therefore five samples ($n_{\textit{meas}}=5$), resulting in a total measurement time of $100.6$ms per microbenchmark. We have manually configured the iteration count ($i = 10^7$ iterations) for the loops in the microbenchmarks to ensure that total execution time always exceeds this duration.

To mitigate external noise and to ensure that the energy measurement are the WCEC, we control several factors that can affect the energy consumption: For the Java runtime system, we disable dynamic class unloading and JIT compilation (executing in interpreted-only mode).
For the execution platform, we fix the CPU frequency to its lowest setting, 1.5GHz, and disable the wireless networking hardware.

\subsection{Benchmark harness}
The benchmark harness consists of 174 microbenchmarks. Each microbenchmark is parameterized by operation, data type, and data size. For the \emph{data types}, we profile 32- and 64-bit integers (\code{int} and \code{long}) and floating-point (\code{float} and \code{double}) data types and reference data types for array and objects. Table~\ref{tab:dataSizes} summarizes the \emph{data sizes} and Table~\ref{tab:ModeledBytecodePatterns} describes all \emph{operations}, where each operation may have a differing amount of Bytecode patterns, e.g., we model three Bytecode patterns for additions on \code{int}, \code{long}, \code{double}, and \code{float} (one for each data size), for a total of 12 addition operation patterns. The array and objects differs from operations on primitive data types. Where we model one array per primitive data type of size one. For objects we model objects from classes and static nested classes, both including four fields, one per primitive data type.  

\begin{table}[H]
    \begin{center}
    \begin{tabular}{l|lllll}
         \toprule
          \textbf{Type} & \textbf{int} & \textbf{float} & \textbf{double} & \textbf{long} & \textbf{reference}\\
         \midrule
         Load & \code{iload} & \code{fload} & \code{dload} & \code{lload} & \code{aload}\\ 
         Constant & \code{iconst} & \code{fconst} & \code{dconst} & \code{lconst} &  \code{aconst}\\ 
         32-bit  & \code{ldc} & \code{ldc} & - & - & -\\
         64-bit  & - & - & \code{ldc2\_w} & \code{ldc2\_w} & - \\
         \bottomrule
    \end{tabular}
    \end{center}
    \caption{Data sizes for the modeled data types, in Bytecode}
    \label{tab:dataSizes}
\end{table}

\section{Evaluation}
\label{sec:evaluation}
To assess our model's accuracy, we examine if the model's predictive distribution of a program fits within the measured energy distribution. 

The model uses Gaussian distributions to describe the energy consumption of source code statements. To model a sequence of statements, we combine several such distributions into a single Gaussian distribution. 

The convolution of independent Gaussian distributions is equivalent to a Gaussian distribution with mean and standard deviation of the sum of the other distributions mean and standard deviation.

Since our prediction uses the distribution from our model, which combines information from a mutual distribution during its creation, we cannot assure that the prediction's distributions are independent. Hence, when considering two or more correlated Gaussian distributions, we also consider their covariance~\cite{SumNormalDistrubutions}: 
\begin{align*}
&\sum^{n}_{i=1}\mathit{Normal}(\mu,\sigma) \sim \mathit{Normal}\left(\sum^{n}_{i=1}\mu_i, \sum^{n}_{i=1}\sigma_i^2\right)\\
&\sigma_i = 
\begin{cases}
      \sqrt{ \sigma_{i-1}^2 + \sigma_{i}^2 + 2 \times \text{cov}(\sigma_{{i-1}}, \sigma_i) } & \text{for } i>1\\
      \sigma_i & \text{for } i=1\
\end{cases}
\end{align*}

As our work defines a methodology to estimate the energy consumption of source code statements, i.e.\@ is not a program analysis tool, we manually simulate how an abstract interpretation-based program analysis could compute energy predictions for various programs and functions, with the convolution of the source code statements' energy distributions.

To assess the validity of our predictions, we compare the model's prediction against energy measurements of $N \times N$ matrix multiplication and calculating $N$ values of the Fibonacci sequence. We chose these two algorithms to assess our model's accuracy when dealing with algorithms primarily using one of the JVM's runtime data areas: heap or local memory (stack).

Our implementation for matrix multiplication utilizes a one dimensional array to represent the matrices (see Fig.~\ref{fig:matrixMult_code}), and only multiplies square matrices that are populated with random values. Whilst the Fibonacci sequence implementation (see Fig.~\ref{fig:Fibonacci_code}), merely iteratively adds values and does not need pre-population.

To assess the model's precision with different data types, we measure the matrix multiplication of \code{int}, \code{long}, \code{float}, and \code{double} matrices. The energy consumption measurements, for both algorithms, use the same methodology as our model's measurements, except that they do not execute in a tight loop of $10^7$ iterations, instead the matrix multiplication executes 100 times and the Fibonacci sequence executes $10^5$ times. 

\begin{figure}
    \centering
    \begin{lstlisting}[language=Java,mathescape]
    matrix1 = $\langle\textsl{NxN 1d-array}\rangle$
    matrix2 = $\langle\textsl{NxN 1d-array}\rangle$
    resultMatrix = $\langle\textsl{NxN 1d-array}\rangle$
    for i = 0 to N
        for j = 0 to N
            for k = 0 to N
                resultMatrix[i*N+j] += matrix1[i*N+k] * matrix2[k*N+j]
    \end{lstlisting} 
    \caption{Pseudo code for matrix multiplication}
    \label{fig:matrixMult_code}
\end{figure}

\begin{figure}
    \centering
    \begin{lstlisting}[language=Java,mathescape]
    f1 = 0, f2 = 1, f3 = 0
	 for i = 0 to N
        f3 = f2 + f1 
        f1 = f2
        f2 = f3
    \end{lstlisting} 
    \caption{Pseudo code for the Fibonacci sequence}
    \label{fig:Fibonacci_code}
\end{figure}

\subsection{Threats to Validity}\label{sec:threats}
The execution platform may produce noise during the benchmark harness's execution due to background processes and other confounding factors. We turned off the platform's wireless networking hardware to mitigate some of these confounds and took multiple samples during varying time points to filter out external noise.

The statistical model's definition, which is the foundation of the research, can have a misguided/illicit construction. We make several assumptions supported by previous research and the distribution's assumption. These actions mitigate subjective biases in the model. However, the choice of model parameters, their distribution, and the parameters themselves are debatable, as with any other model.
Our justifications for the model parameters stem from the JVM specification and programming languages' definition, where we model aspects parametrized within Java Bytecode. 
Using the Gaussian distribution is justified by ontological and epistemological aspects \cite{rethinking}. The ontological reasoning is that the Gaussian distribution regularly appears when modelling nature. Occurring since Gaussian distribution models processes that sum fluctuations, and according to the central limit theorem, thus converging to a standard normal distribution. 

The epistemological reasoning is that we do not know much about the measurements other than that they have a finite mean and variance. Hence, it represents our uncertainty, since these are the two aspects of the energy consumption behaviour we allow ourselves to assume.

Other threats mostly pertain to the execution platform and JVM.
The memory state before each bytecode pattern execution could alter the result.
Hence, we execute the garbage collector before each measurement cycle.
To mitigate bias due to paging, caching, and disruptions from concurrent processes we measure each bytecode pattern ten times, using a split-plot approach, again randomizing execution order across sequences.

To mitigate energy differences due to dynamic frequency scaling and to obtain the device's WCEC we set the CPU to its lowest frequency of 1.5GHz.

\section{Results \& Analysis}
\label{sec:results}
In this section, we present the energy impact of the operation, data size, data type, and device, and the model's accuracy and predictive accuracy. Section~\ref{sec:results:modelAccuracy} describes the model's ability to fit measurements. Section~\ref{sec:results:dataDeviceCategory} describes the energy consumption contribution of the data type, data size, and device, whilst Section~\ref{sec:results:operationCategory} describes the operations' energy consumption. Section~\ref{sec:results:categoryImpact} describes each model category's impact on predictions, and Section~\ref{sec:results:prediction} evaluates our model's predictive accuracy.

\subsection{Model Accuracy}
\label{sec:results:modelAccuracy}
The model uses Markov Chain Monte Carlo (MCMC) to estimate the energy consumption of the programming language. MCMC, by definition, draws samples from the target distribution only after convergence. Hence, it is vital to reach convergence to have a sound model. We check this using two procedures: the $\hat{R}$ statistic and the MCMC's diagnostics. When all chains have reached convergence, $\hat{R} = 1$. $\hat{R}$ value exceeding 1.01 indicates possible convergence issues~\cite{BayesianWorkflow, BayesianEstimationCambridge, BayesianArchive}. To further ensure a reliable model, we look at the effective sample size (ESS), describing the number of independent draws from the posterior distribution. An ESS greater than 400 suggests that the $\hat{R}$ statistic is valid ~\cite{BayesianArchive}, and an $\frac{ESS}{N} > 0.0001$ suggests a trustworthy ESS estimator~ \cite{BayesianGit}. A model with $\hat{R} < 1.01$ and $ESS > 400$ suggests that the sampler performs well~\cite{BayesianArchive}. Our model's parameters have a $\hat{R}$ between $(0.999, 1.002)$, where all $ESS > 400$, all $\frac{ESS}{N} > 0.0001$ (see Appendix~\ref{sec:appendix:a}) and no divergences are found by the MCMC diagnostics, suggesting a valid model.  

However, we must also ensure that the model fits the data. We show this with posterior predictive checks. With our model we can depict how accurate our estimation is per Bytecode pattern we model. Our model fails to accurately estimate eight out of our 348 microbenchmark measurements (174 microbenchmarks on two devices). These are array allocations for \code{float}, \code{long} on device 2, array allocation for references on both devices, if a reference is not null on device 1, if a reference is unequal for device 2, and if an \code{int} is equal to another \code{int} on both devices. All other estimations fit the data. The last metric to ensure a good fit for our model is the Monte Carlo Standard Error (MCSE), reflecting the uncertainty of estimating the posterior mean~\cite{BayesianEstimationCambridge}, hence, it should be as low as possible. Note that MCSE in Bayesian inference is not equivalent to the Frequentist concept of standard error, referring to the variance across an infinite number of repeated samples. Our MCSEs are sufficiently small  ($(2.5*10^{-12}\%, 2.1*10^{-8}\%)$ difference from mean) not to influence the estimations (see Appendix~\ref{sec:appendix:a}) greatly.

\subsection{Energy of Data and Device Categories}
\label{sec:results:dataDeviceCategory}

\begin{figure}
    \centering
    \subfloat[\centering Data size]{{\includegraphics[width=.48\linewidth]{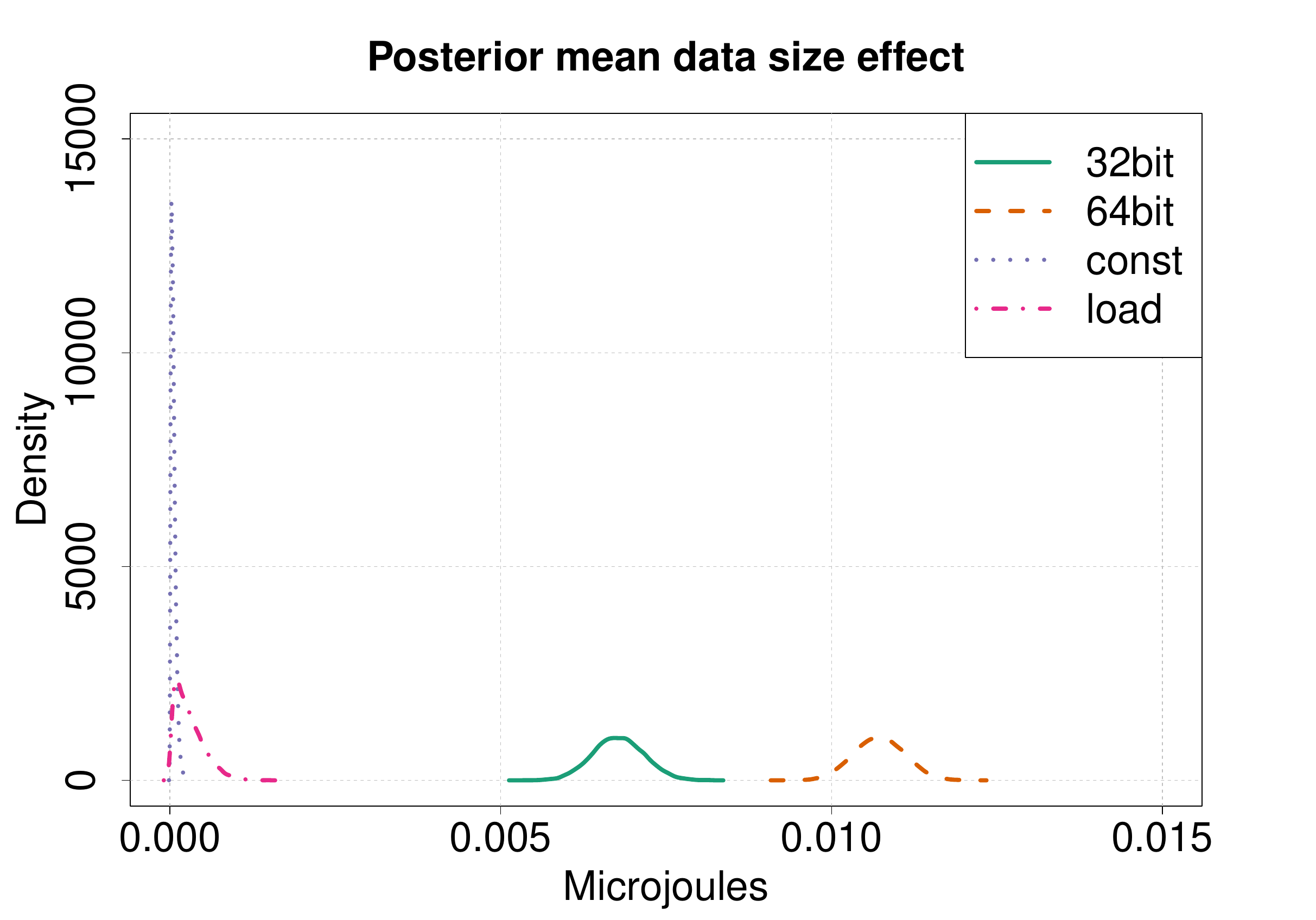} }
    \label{fig:ModelRes_datasize}
    }%
    \subfloat[\centering Data type]{{\includegraphics[width=.48\linewidth]{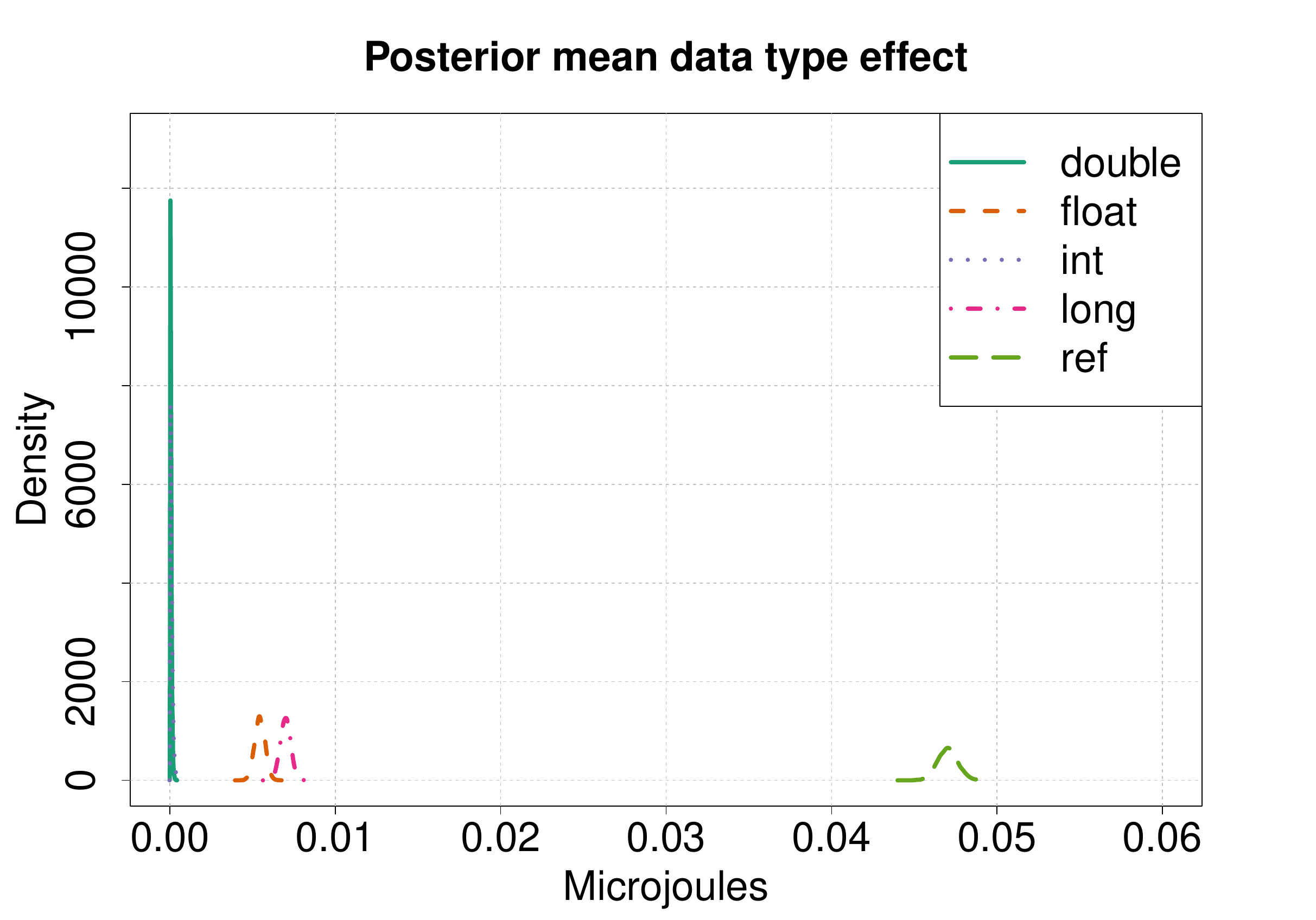} }
    \label{fig:ModelRes_datatype}
    }
    
    \subfloat[\centering Device instance]{{\includegraphics[width=.48\linewidth]{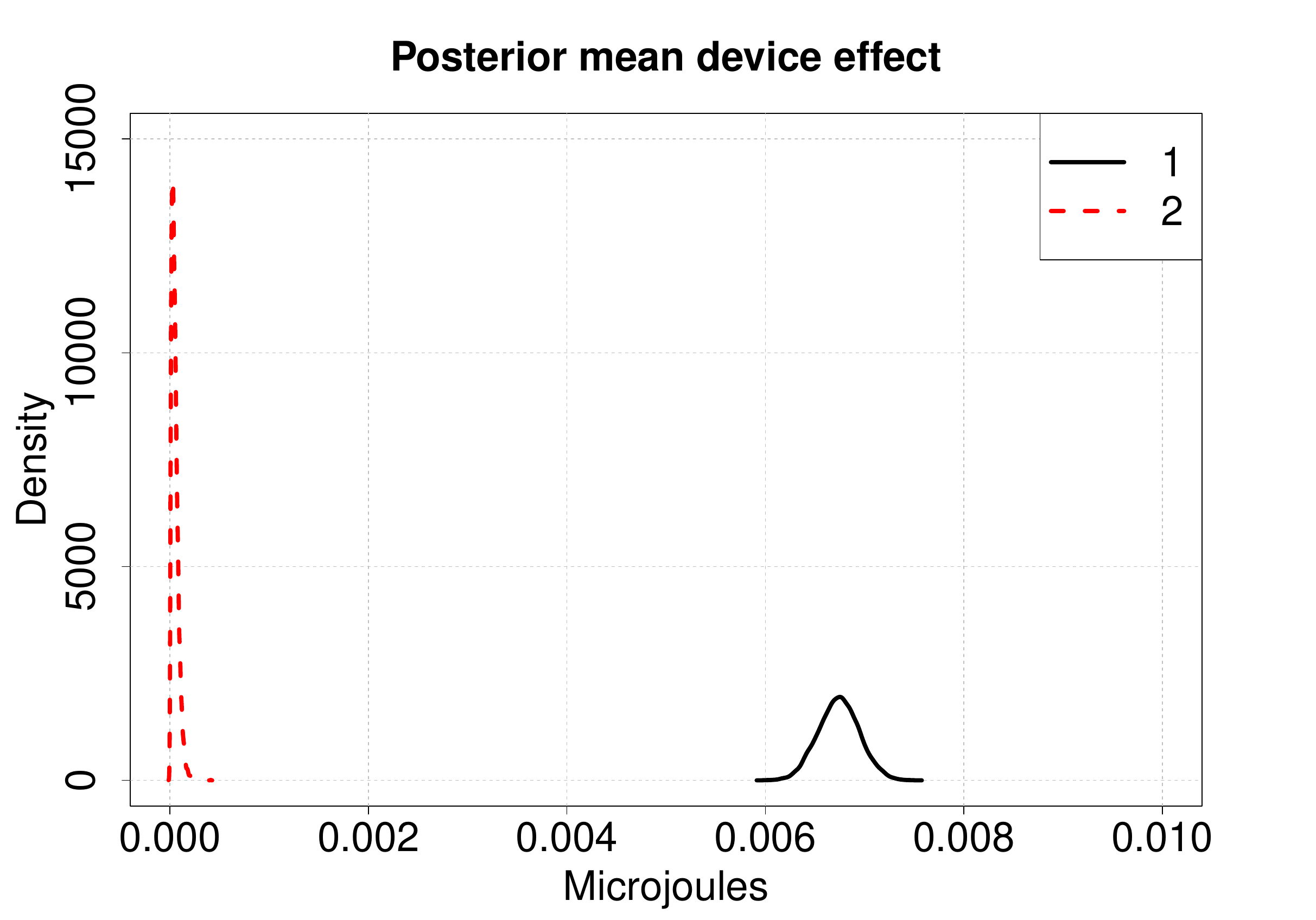} }
    \label{fig:ModelRes_device}
    }%
    \caption{Mean energy distribution of the energy consumption for each category within the model (note the difference in the X and Y-axis)}
\end{figure}

The model shows that for the \emph{data size category} (see Fig.~\ref{fig:ModelRes_datasize}), loading a variable's value from the constant pool and declaring a variable with a constant value consumes the least energy.
Unsurprisingly, larger (64-bit) data sizes consume more energy than the smaller (32-bit) ones.
For the \emph{data type} category (see Fig.~\ref{fig:ModelRes_datatype}), using a reference (for an array or an object) consumes considerably more energy than the primitive data types, in which \code{long} consumes the most, followed by \code{float}, \code{int}, and \code{double}. However, the operations that pertain to primitive data types and operations for references are disjoint.
The last category, \emph{device}, (see Fig.~\ref{fig:ModelRes_device}) shows that one device consumes substantially more energy than its functionally identical counterpart.

\subsection{Energy per Operation Category}
\label{sec:results:operationCategory}

\begin{figure}
    \centering
    \subfloat[\centering Arithmetics]{{\includegraphics[width=.48\linewidth]{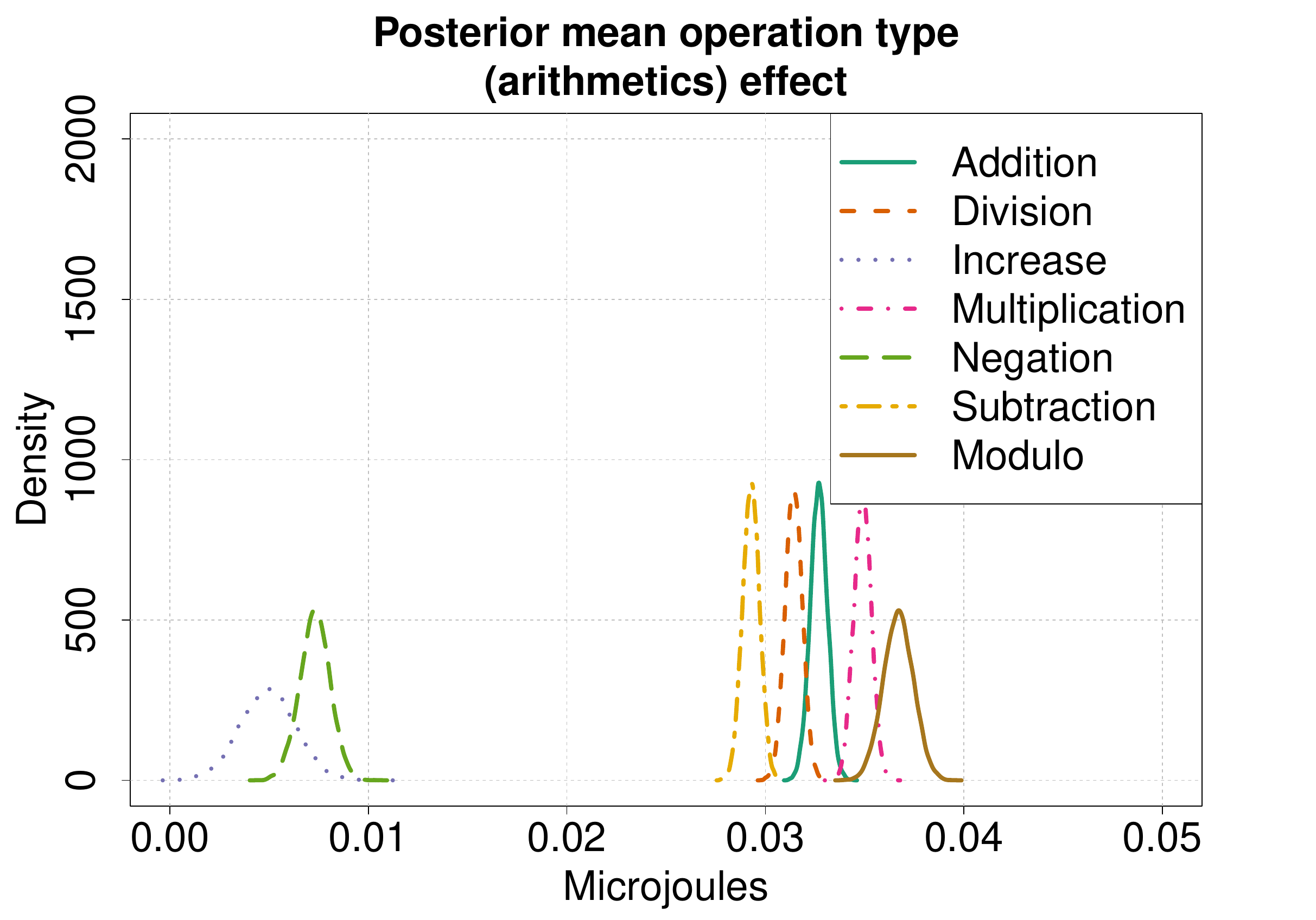} }
    \label{fig:ModelRes_op_arith}
    }%
    \subfloat[\centering Bit Operations]{{\includegraphics[width=.48\linewidth]{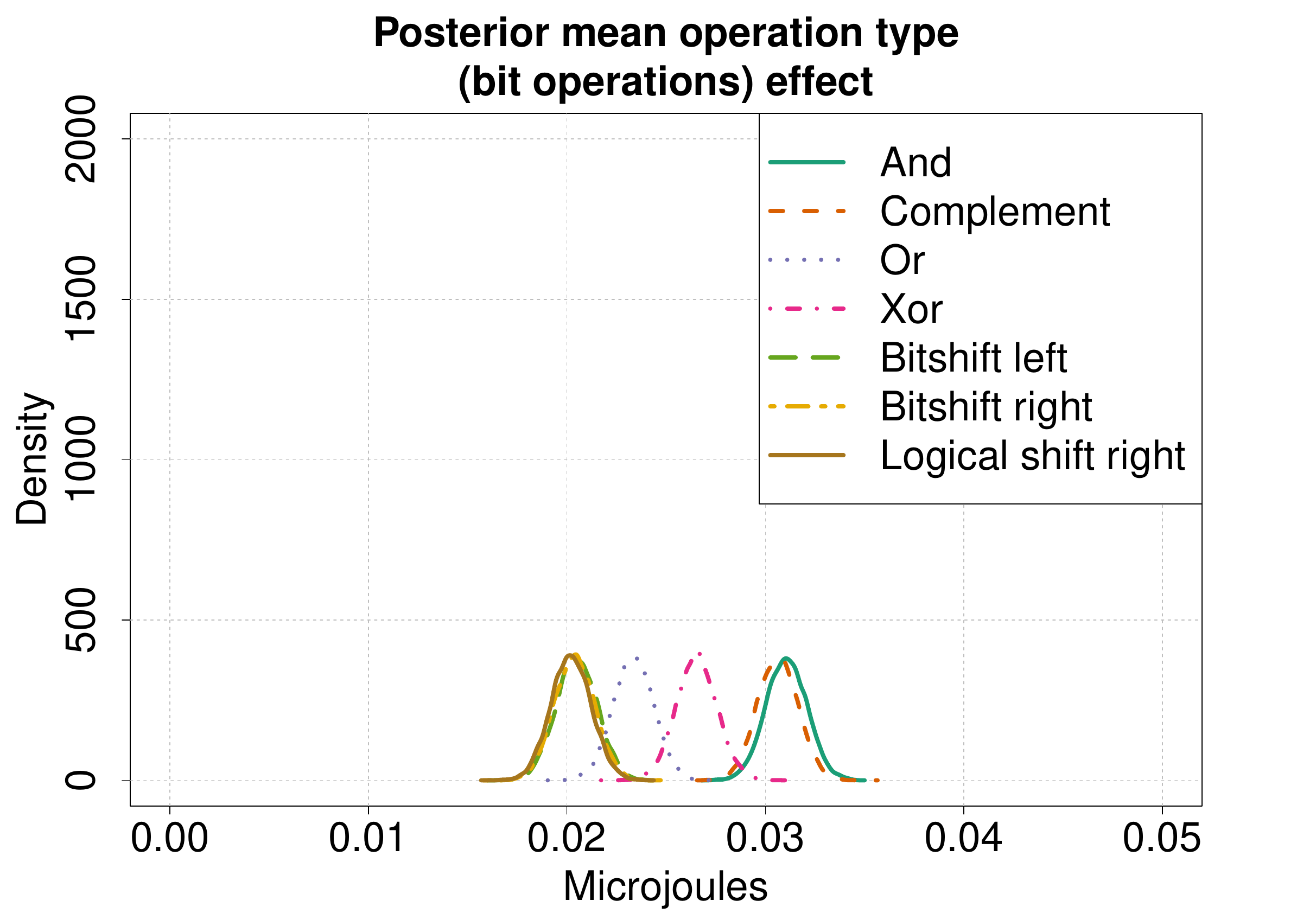} }
    \label{fig:ModelRes_op_bitOps}
    }
    \caption{Mean energy distribution of Arithmetic and Bit operations}
    \label{fig:ModelRes_op_arithbitOps}
\end{figure}

The energy per operation demonstrates the energy consumption for a given operation averaged over all its permissible data types. Hence, it represents energy consumption at runtime.
For \emph{arithmetic and bit operations} (see Fig.~\ref{fig:ModelRes_op_arithbitOps}), most operations consume approximately 0.03 -- 0.04 $\mu$J, where arithmetic operations generally consume more than bit operations. Two operations break the trend: the increase and negation operations. The negation operation negates a value and is defined for all primitive data types in the study. While the increment operation increments a value by a constant, it is only applicable for integer data types.  

\begin{figure}
    \centering
    \subfloat[\centering \code{float} and \code{double} conversion]{{\includegraphics[width=.48\linewidth]{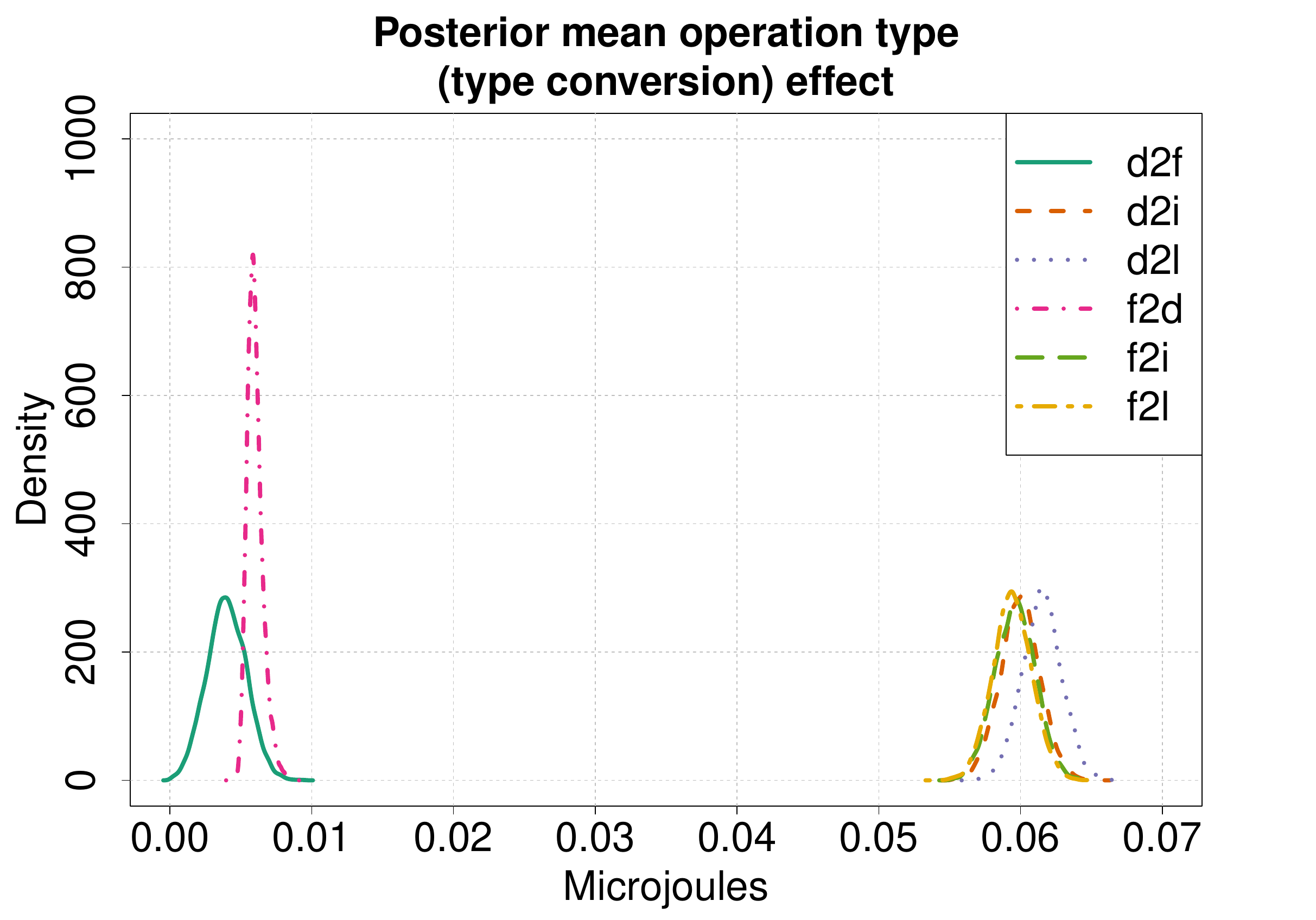} }
    \label{fig:ModelRes_op_typeConvFloat}
    }%
    \subfloat[\centering \code{int} conversion]{{\includegraphics[width=.48\linewidth]{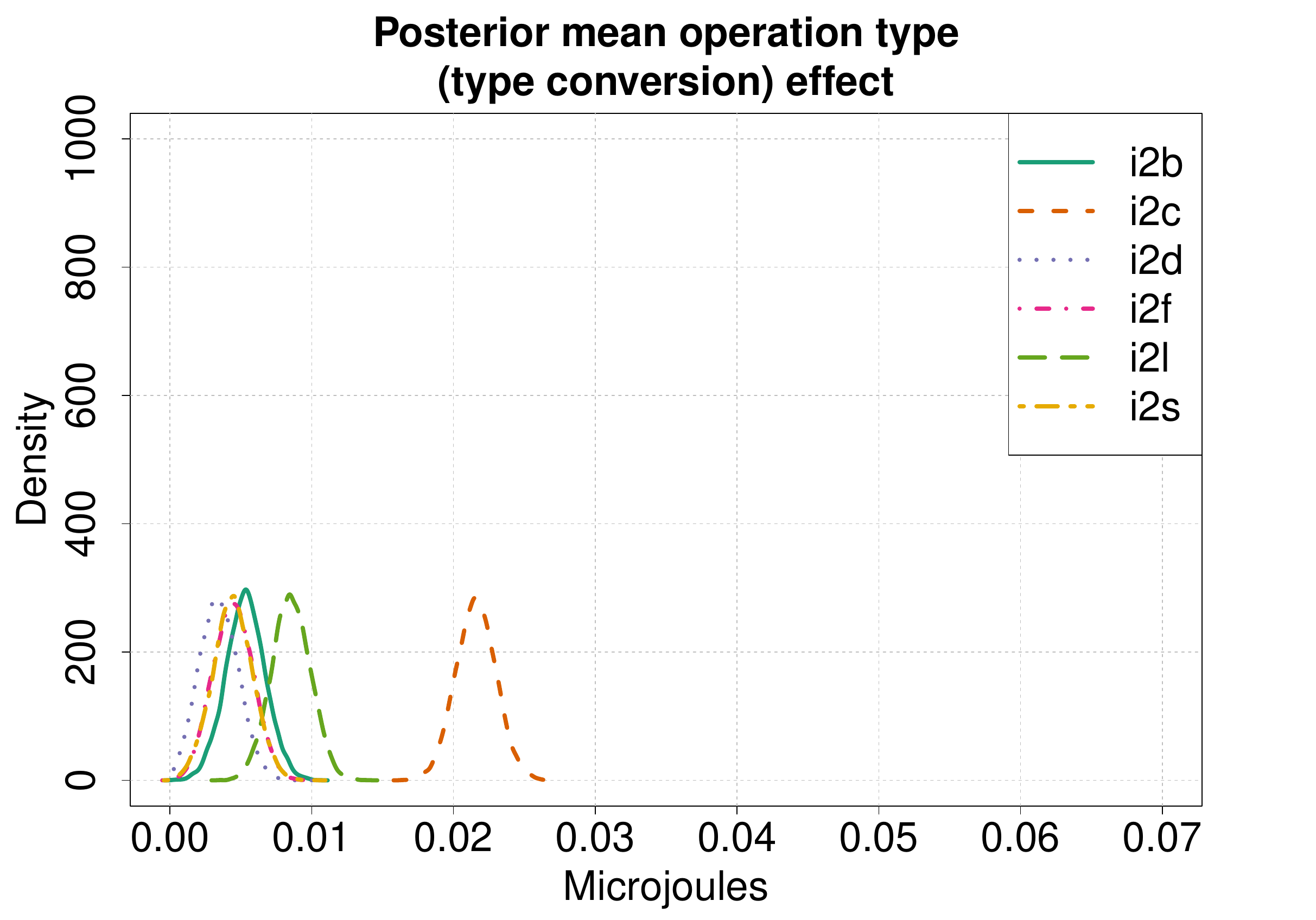} }
    \label{fig:ModelRes_op_typeConvInt}
    }
    
    \subfloat[\centering \code{long} conversion]{{\includegraphics[width=.48\linewidth]{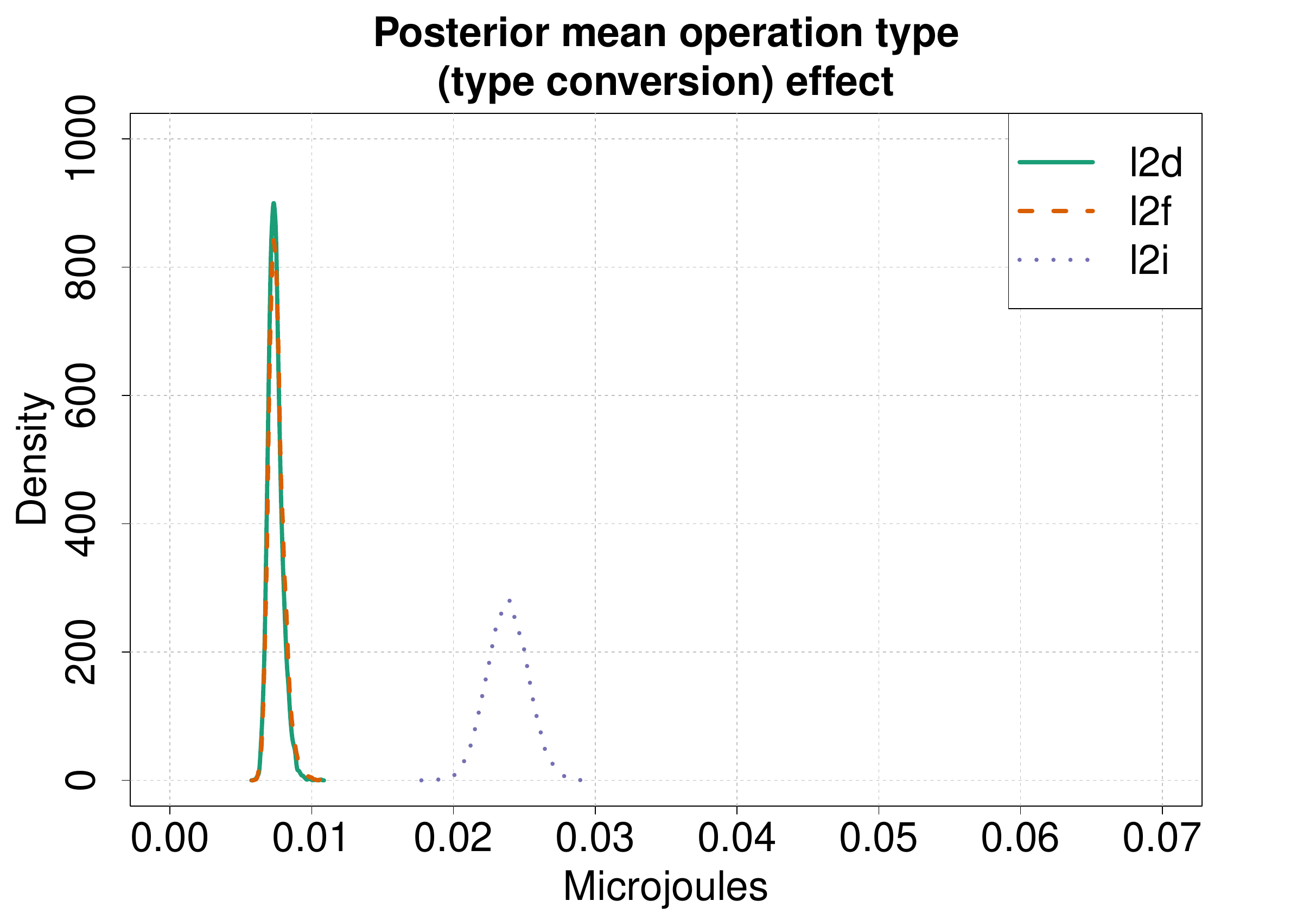} }
    \label{fig:ModelRes_op_typeConvLong}
    }%
    \caption{Mean energy distribution of type conversion operations}
    \label{fig:ModelRes_op_typeConv}
\end{figure}

\emph{Type Conversion} (see Fig.~\ref{fig:ModelRes_op_typeConv}) consumes the least energy when widening an integer variable to a floating-point variable, widening a 32-bit variable to 64-bit, or narrowing a \code{double} to a \code{float}. In contrast, the most energy-consuming operation is narrowing a floating-point variable to an integer, where narrowing a \code{double} requires more energy than narrowing a \code{float}. Generally, widening a variable tends to consume less energy than narrowing one.

\begin{figure}
    \centering
    \subfloat[\centering Array operations]{{\includegraphics[width=.48\linewidth]{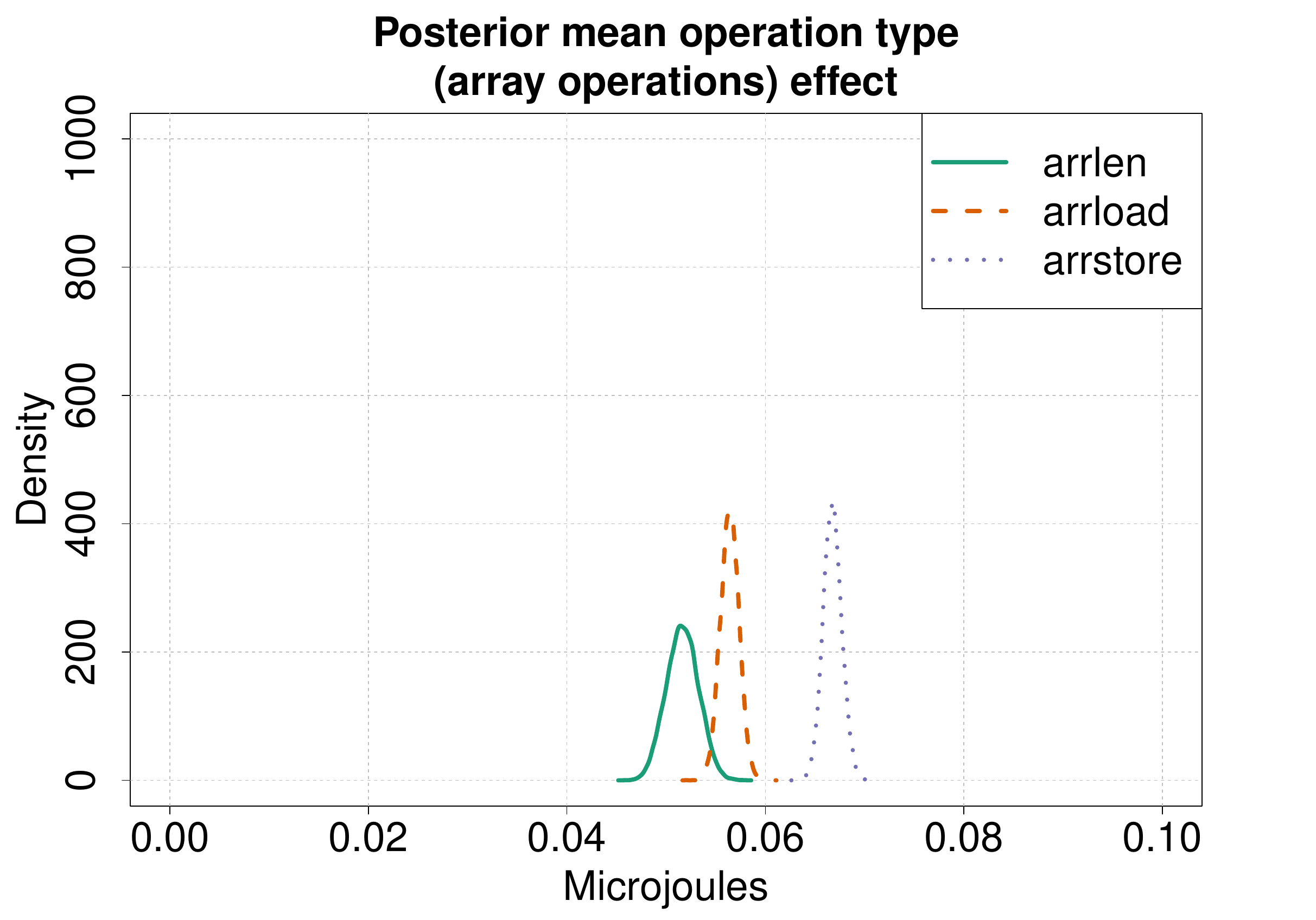} }
    \label{fig:ModelRes_op_arrOps}
    }%
    \subfloat[\centering Operations on objects]{{\includegraphics[width=.48\linewidth]{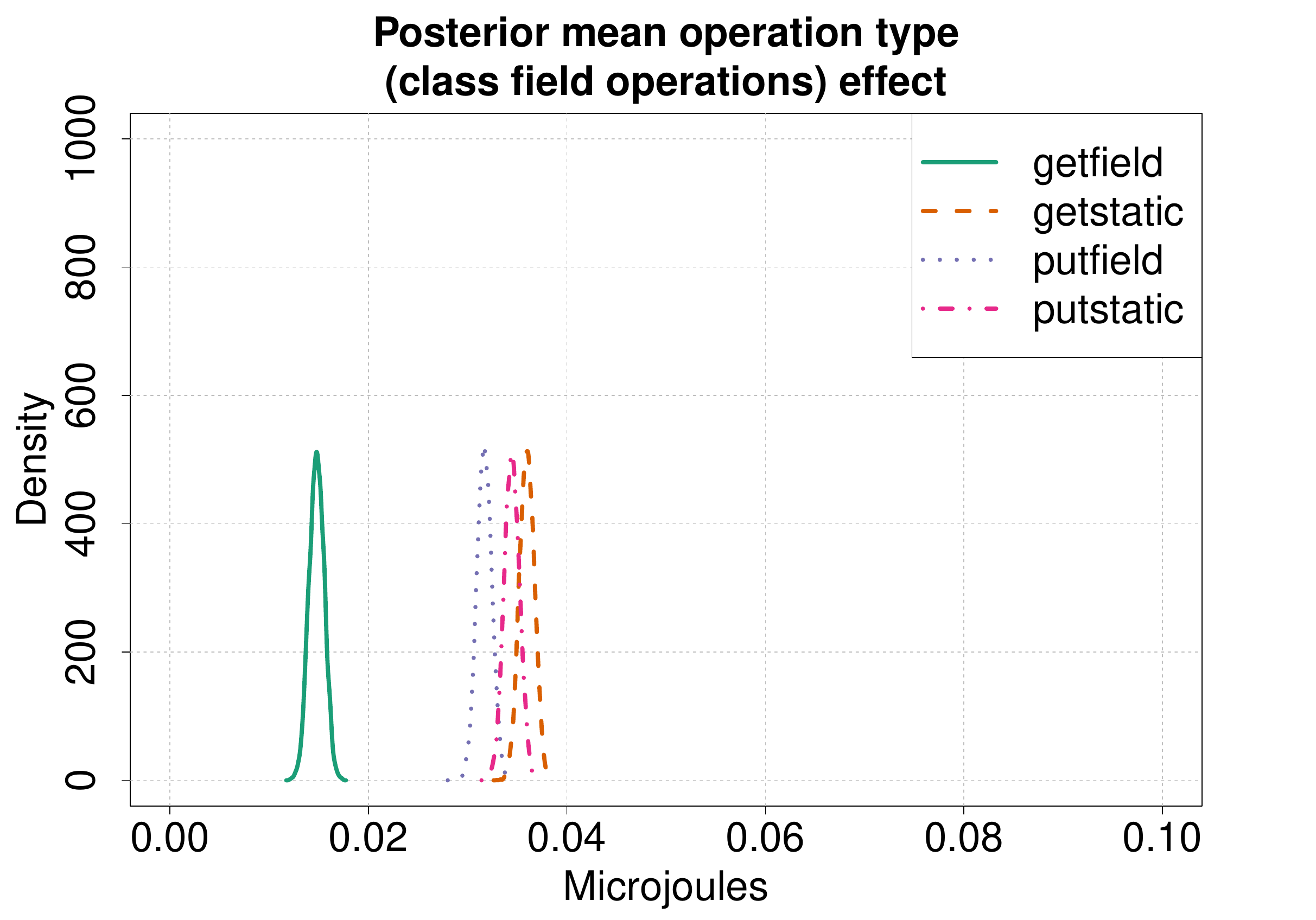} }
    \label{fig:ModelRes_op_classOps}
    }
    
    \subfloat[\centering Variable and object creation]{{\includegraphics[width=.48\linewidth]{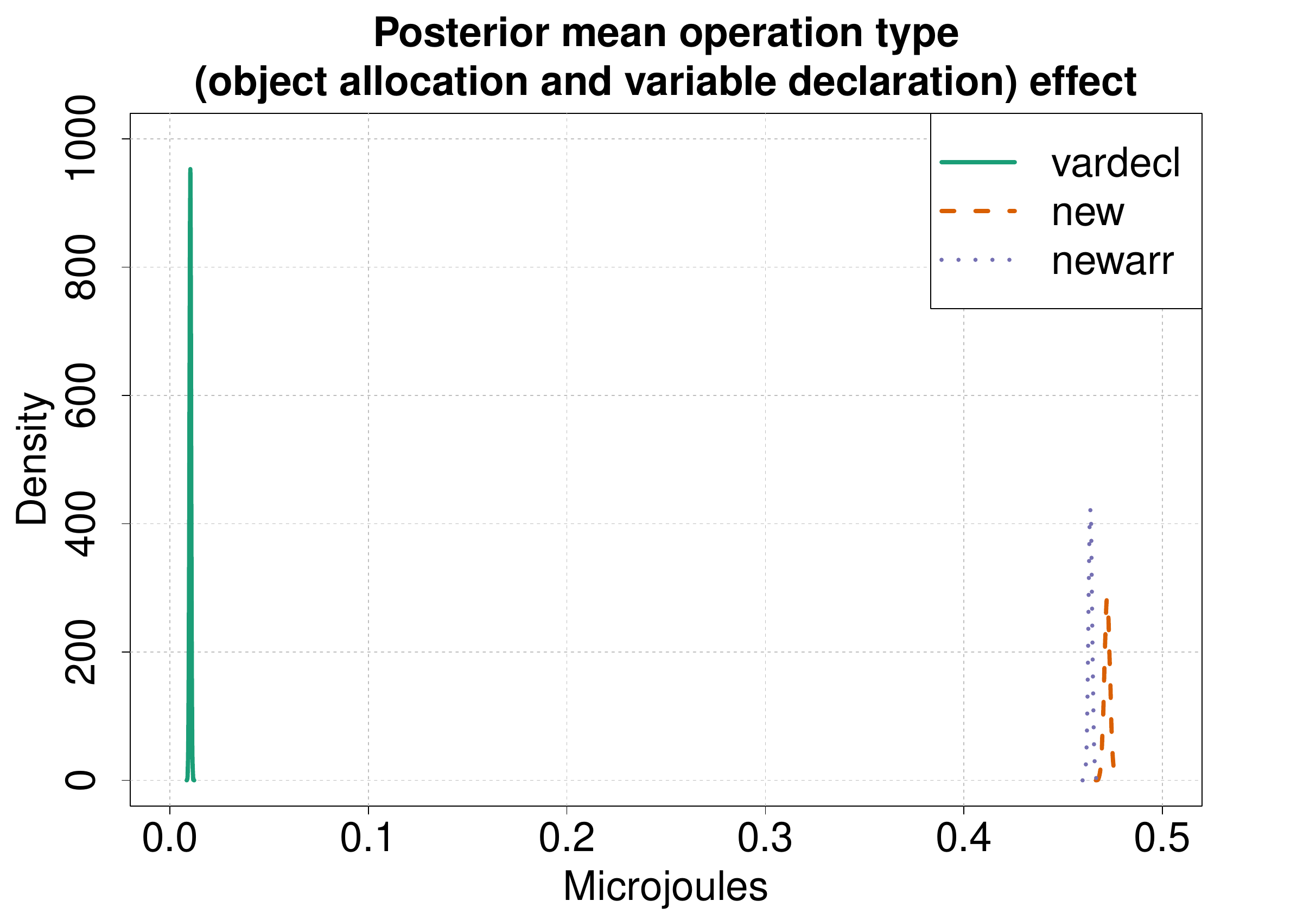} }
    \label{fig:ModelRes_op_objCreation}
    }%
    \subfloat[\centering Method invocation and return]{{\includegraphics[width=.48\linewidth]{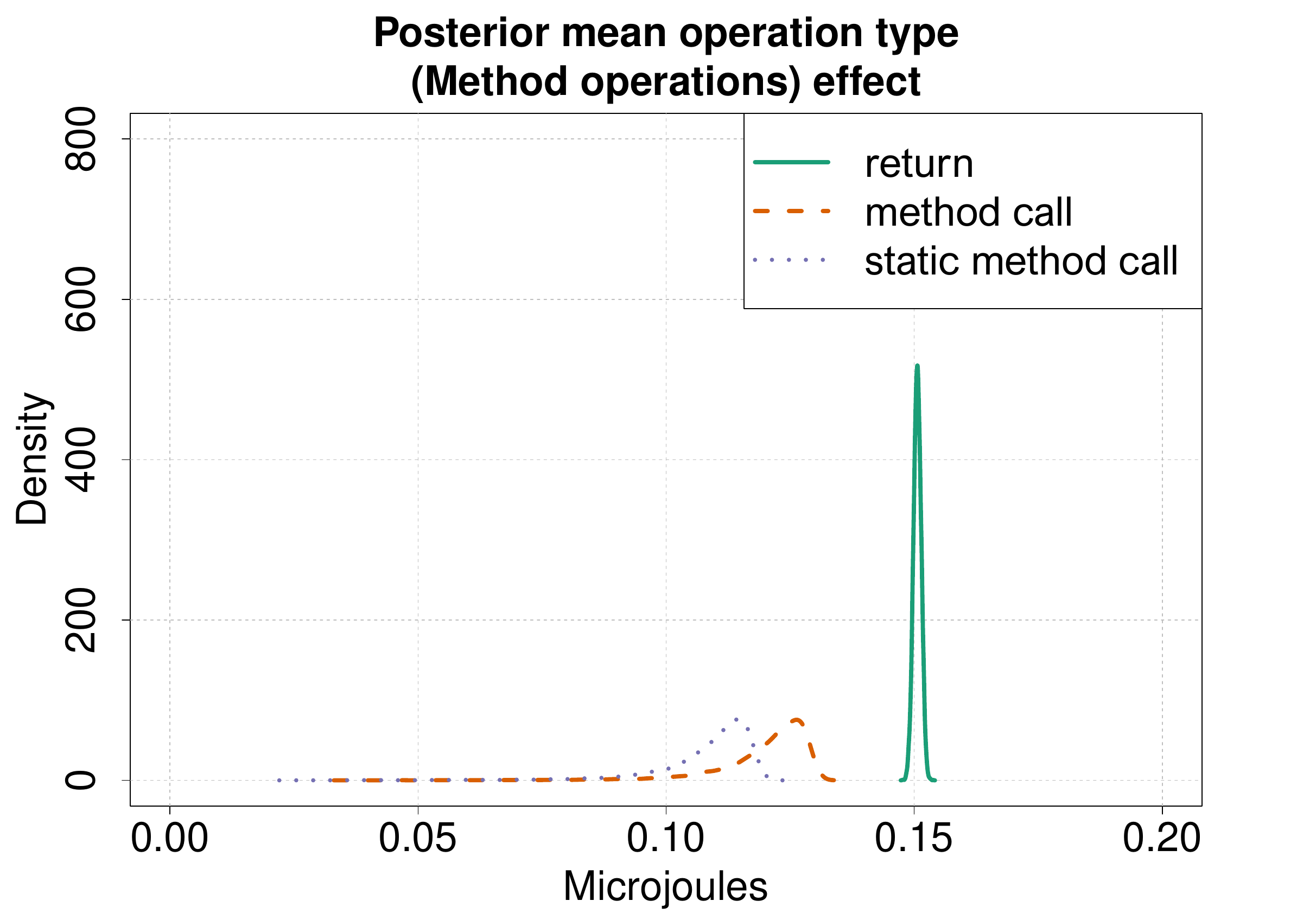} }
    \label{fig:ModelRes_op_method}
    }%
    \caption{Mean energy distribution for declarations, array allocation and accesses, object allocations, field accesses, and method invocations (note the X-axis difference in Fig~\ref{fig:ModelRes_op_objCreation} \& \ref{fig:ModelRes_op_method})}
    \label{fig:ModelRes_op_objectOps}
\end{figure}

\emph{Allocations, Method, Class, and Array operations} includes the most energy-consuming operation in the study, dynamic allocation of an object or array (see Fig.~\ref{fig:ModelRes_op_objectOps}). 
The consumption when manipulating an object's field variables differs depending on whether it is a static field, where static ones consume more energy for both reads and writes (see Fig.~\ref{fig:ModelRes_op_classOps}). 
Reading consumes less than writing, which also holds for array read/write (see Fig.~\ref{fig:ModelRes_op_arrOps}).  
The static consumption trend continues for method calls but in reverse (see Fig.~\ref{fig:ModelRes_op_method}), where the non-static method calls consume more energy than static ones. However, returning a variable consumes more energy than the method calls.

\begin{figure}
    \centering
    \subfloat[\centering If-statements for \code{int}, $\neq0$ ]{{\includegraphics[width=.48\linewidth]{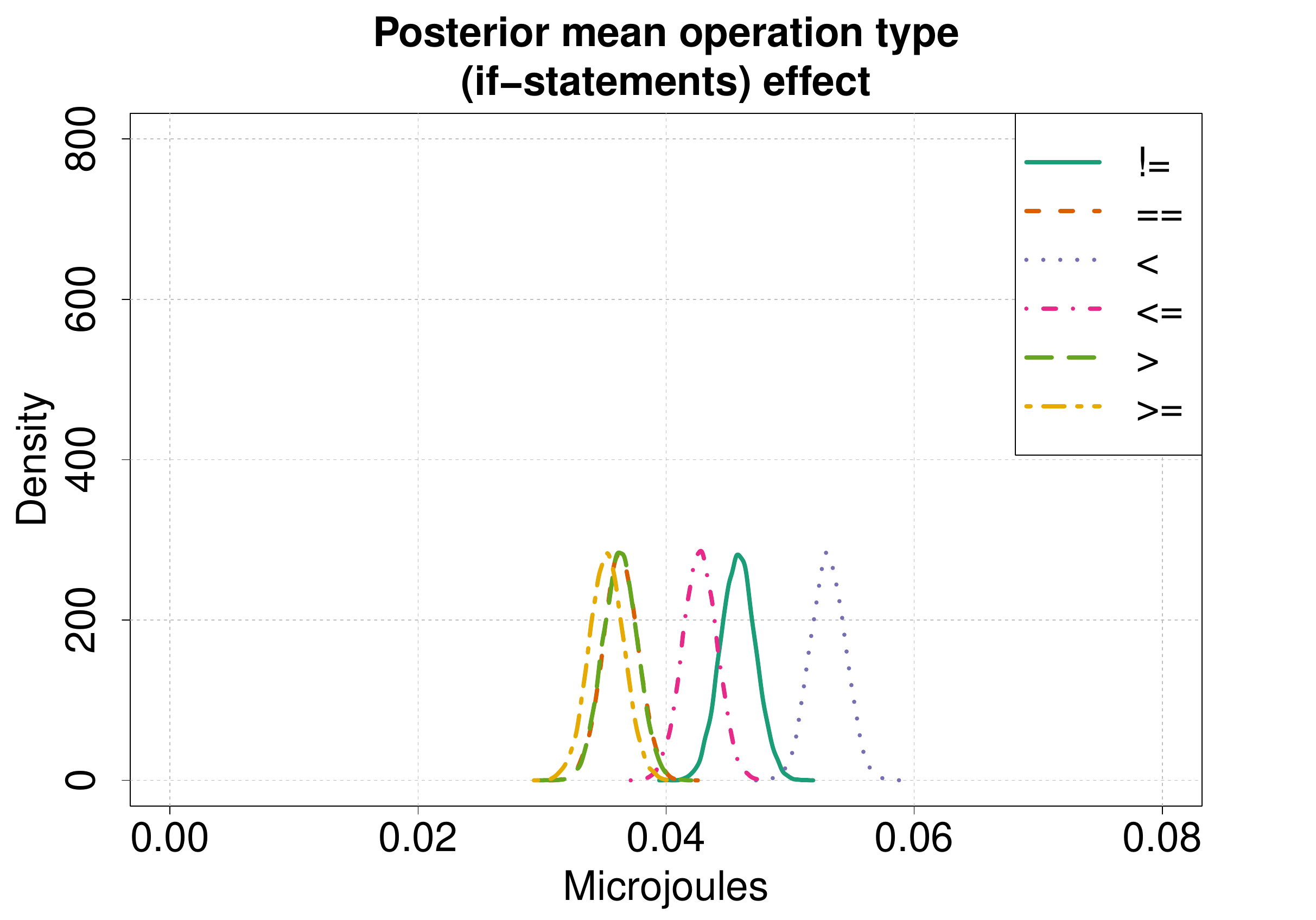} }
    \label{fig:ModelRes_op_if_int}
    }%
    \subfloat[\centering If-statements for numerical data types and \code{int} $= 0$]{{\includegraphics[width=.48\linewidth]{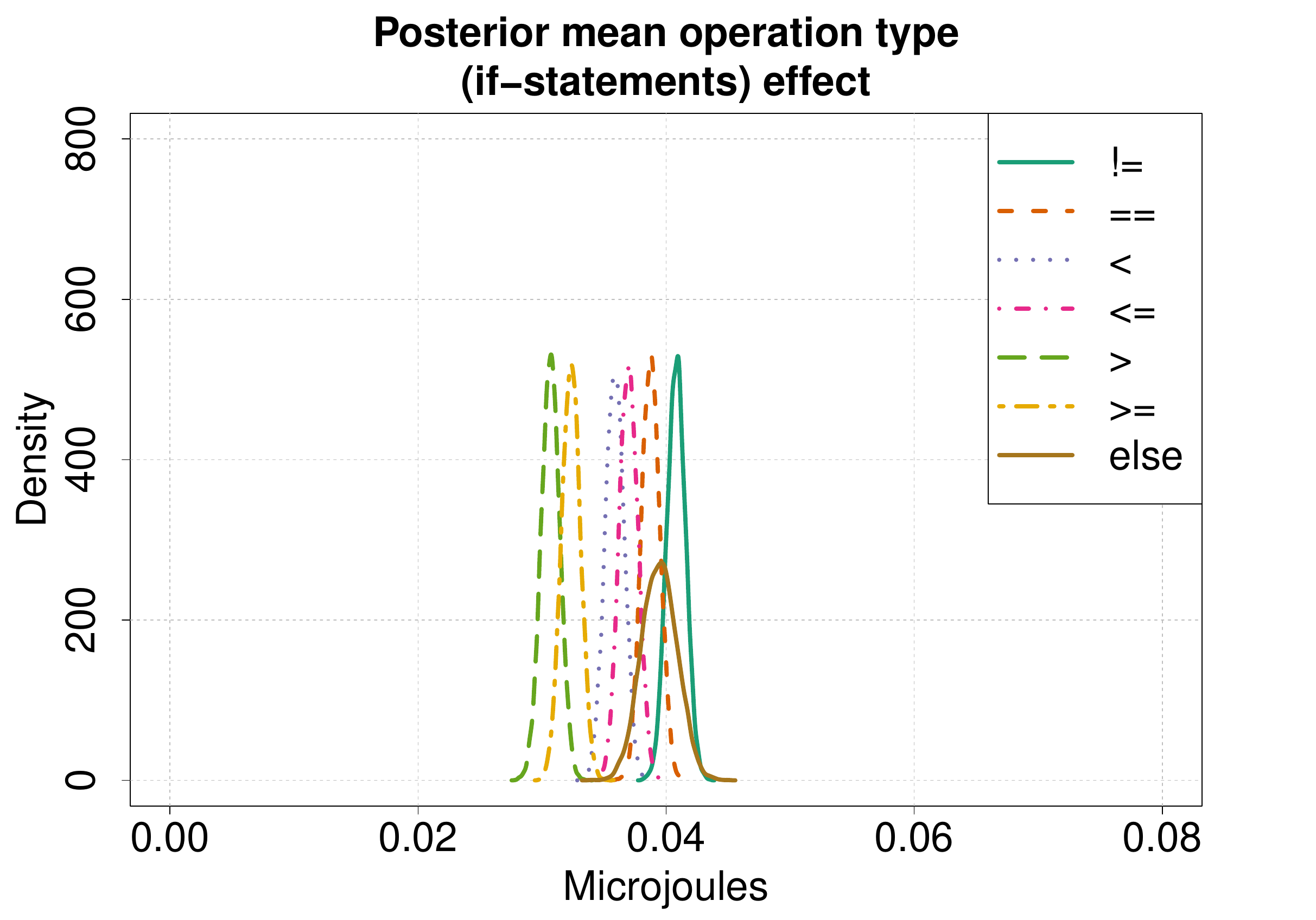} }
    \label{fig:ModelRes_op_if_all}
    }
    
    \subfloat[\centering If-statements for references]{{\includegraphics[width=.48\linewidth]{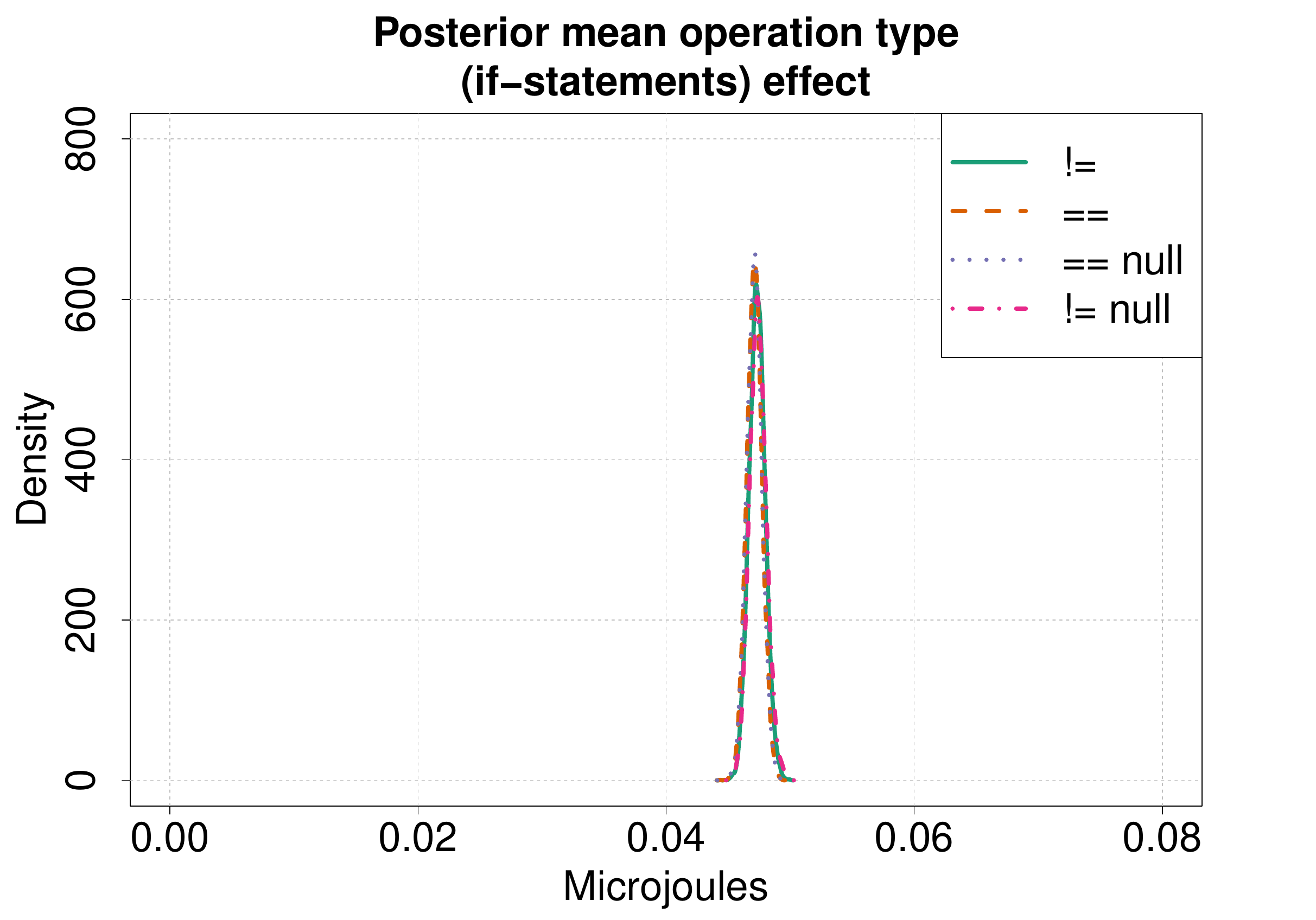} }
    \label{fig:ModelRes_op_if_ref}
    }%
    \subfloat[\centering Switch case]{{\includegraphics[width=.48\linewidth]{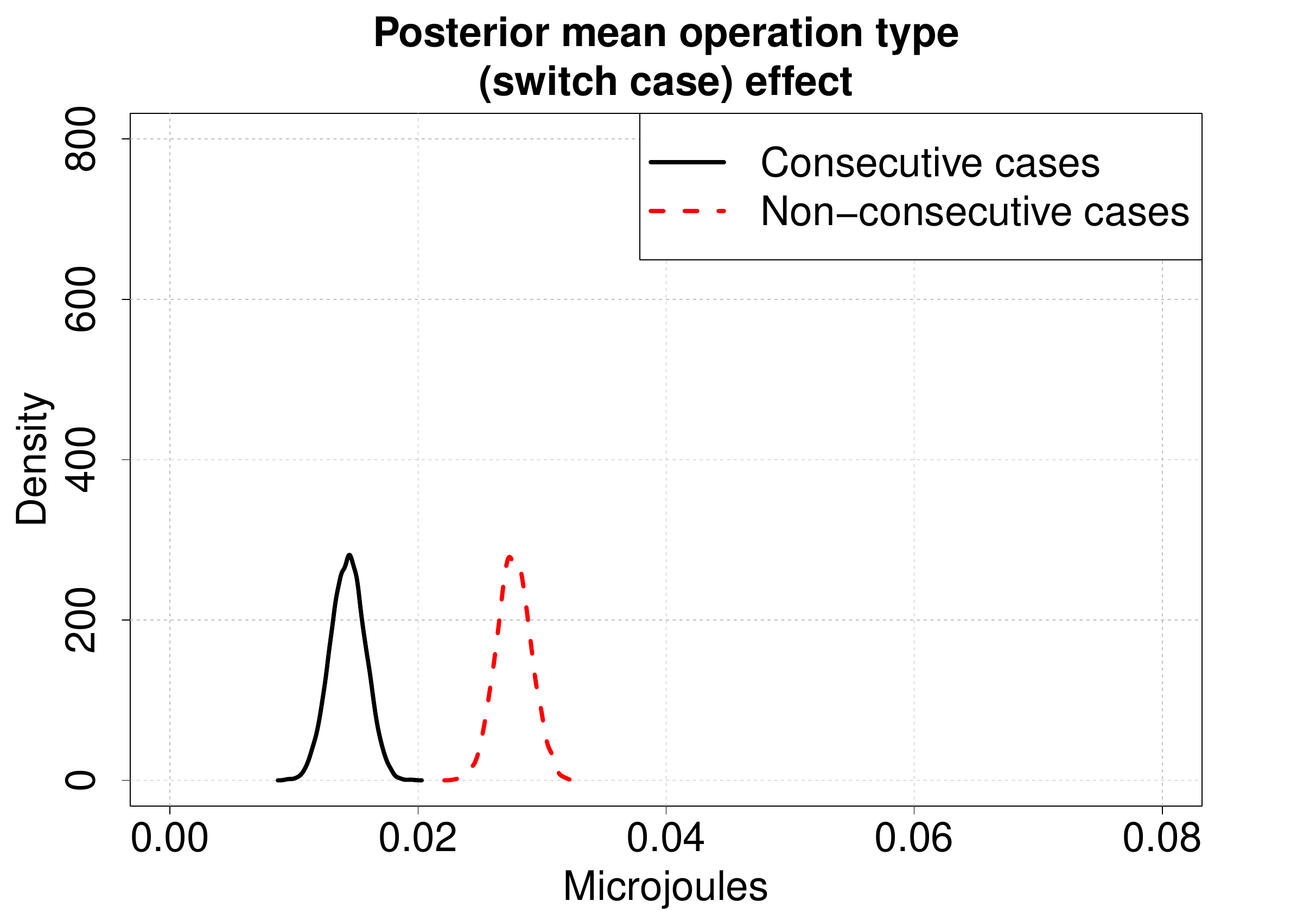} }
    \label{fig:ModelRes_op_switch}
    }%
    \caption{Mean energy distribution for If statements}
    \label{fig:ModelRes_op_if}
\end{figure}

\emph{Control transfer} encompasses several Bytecode operations.
Java Bytecode uses three different operations depending on the data type in the if-statement. When evaluating an \code{int}, $\neq 0$ (see Fig. \ref{fig:ModelRes_op_if_int}) it uses the \texttt{if\_icmp<cond>} operation.  When evaluating object or array references (see Fig. \ref{fig:ModelRes_op_if_ref}) it uses \texttt{if\_acmp<cond>}, and for all the other primitive data types and for \code{int}, $= 0$ (see Fig. \ref{fig:ModelRes_op_if_all}) Java Bytecode uses \texttt{if<cond>}.
When comparing numerical data types, the energy consumption is approximately equivalent, whether the \code{int} is zero or not.
Energy consumption for if-statements that evaluate reference types does not vary substantially.
The switch-cases only apply to \code{int}, and have two Java Bytecode operations (see Fig. \ref{fig:ModelRes_op_switch}), depending on their cases. When the cases are in consecutive order, Java Bytecode uses \texttt{tableswitch} and otherwise the \texttt{lookupswitch} operation, where the former is the least consuming branching operation in the study. 

\subsection{Energy Categories Impact}
\label{sec:results:categoryImpact}
\begin{figure}
    \centering
    \includegraphics[width=.8\linewidth]{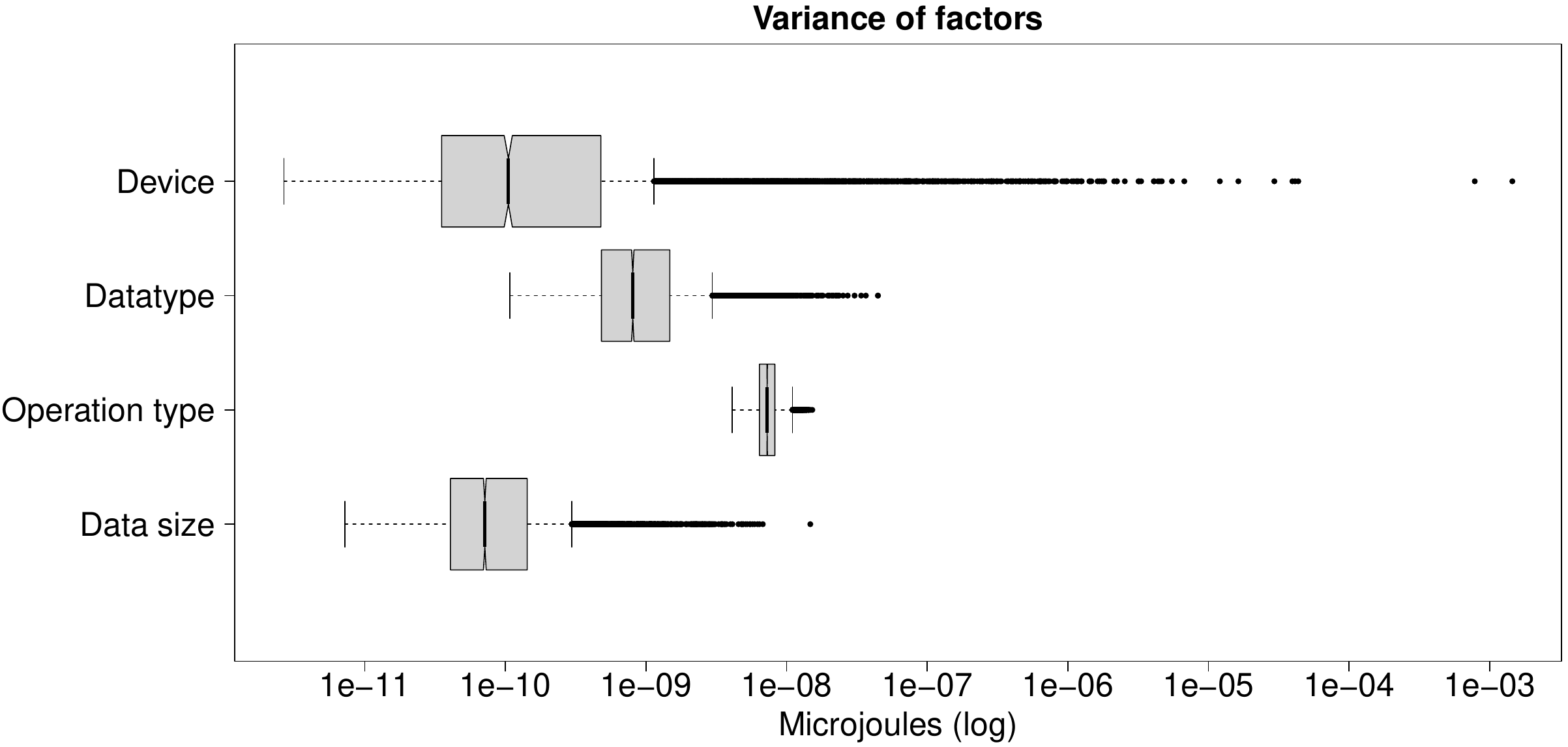}
    \caption{Variance of the model's categories, larger values account for larger variation in the model}
    \label{fig:variance}
\end{figure}

The four categories (operation, data type, data size, and device) in the model differ in their impact on the final energy consumption prediction (see Fig.~\ref{fig:variance}). A prediction is a Gaussian distribution, where its mean is composed of a linear combination of the elements from the quadruple $F$ (Section \ref{sec:Method:Meas:BenchmarkHarness}), $\textit{Normal}(o+t+s+d, \sigma)$, where the Bayesian model estimates $\sigma$. For example, an \code{int} array allocation has the following quadruple describing its bytecode pattern and device's energy consumption: $F_x = \left<\textit{newarray}, \textit{int}, \textit{const}, \textit{device 1}\right>$. Hence, the prediction for this operation will be: $\textit{Normal}(\sum F_i, \sigma)$. Each element in the quadruple affects the energy consumption to a different extent. Our model shows that the operation type has the most substantial effect on energy consumption, followed by the data type, while the data size and device instance have approximately the same effect.

\subsection{Predictions}
\label{sec:results:prediction}

Hao et al.'s eCalc~\cite{ecalc} predicted the energy consumption of programs by summing the mean energy consumption of each instruction in an execution trace.
Here, we adapt their approach, using convolution for the energy consumption distributions of all source code statements for matrix multiplication and the Fibonacci sequence.
Since we sum Gaussian distributions, our mean and standard deviation increase per source code statement in the prediction.

\begin{figure}
    \centering
    \subfloat[\centering Matrix Multiplication \code{int}]{{\includegraphics[width=.48\linewidth]{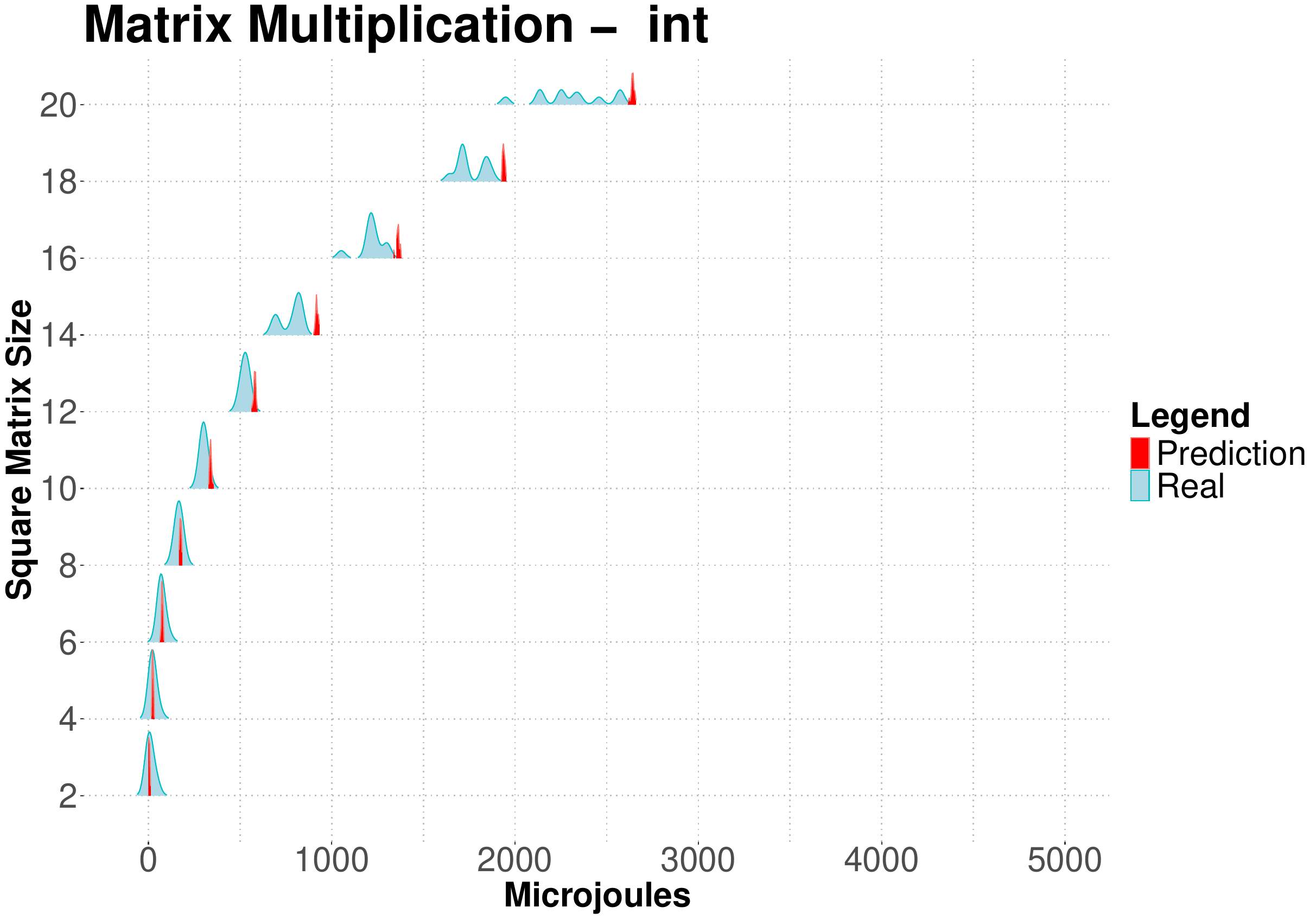} }
    \label{fig:predInt}
    }%
    \subfloat[\centering Matrix Multiplication \code{long}]{{\includegraphics[width=.48\linewidth]{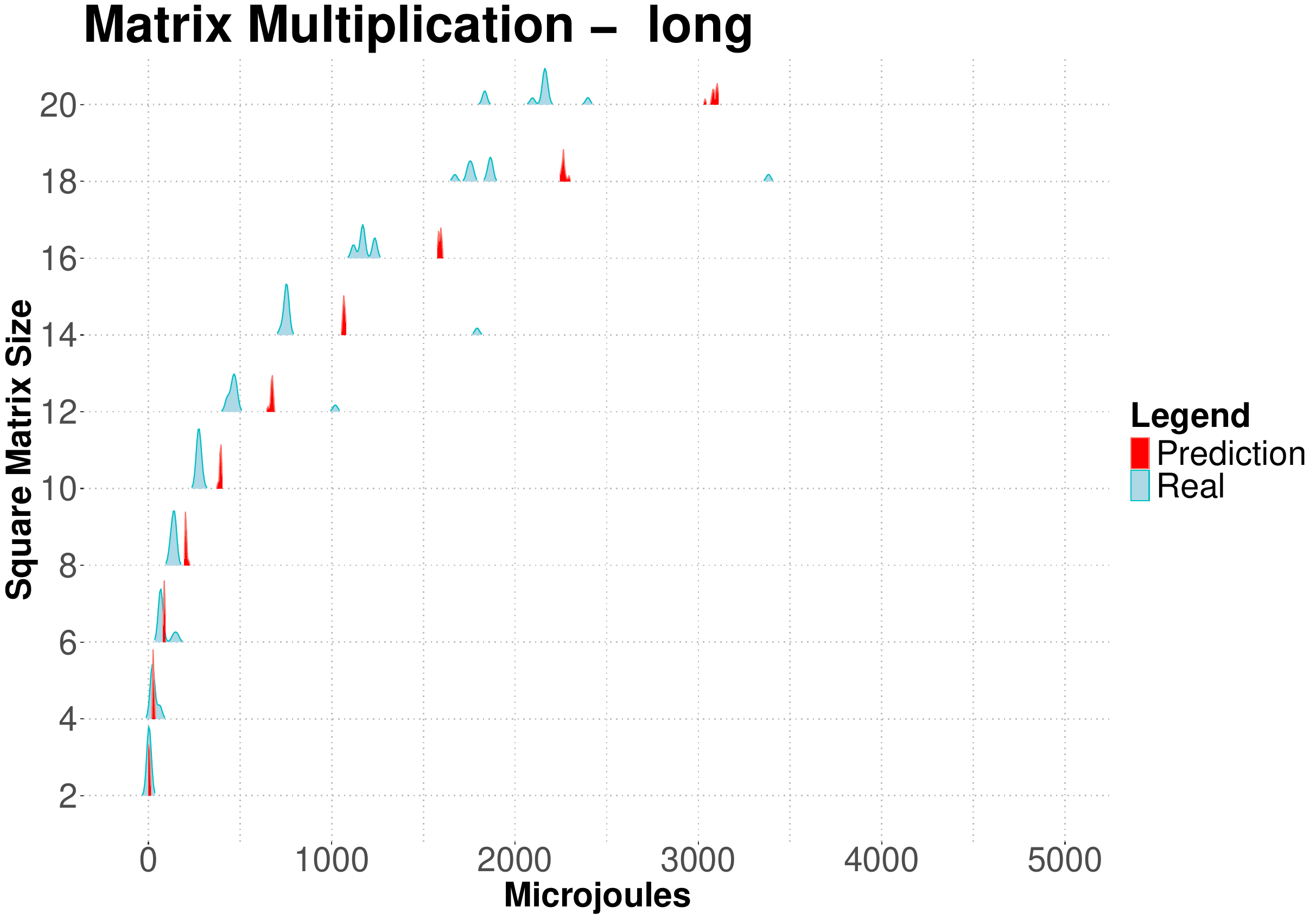} }
    \label{fig:predLong}
    }%
    \vspace{1cm}
    \subfloat[\centering Matrix Multiplication \code{double}]{{\includegraphics[width=.48\linewidth]{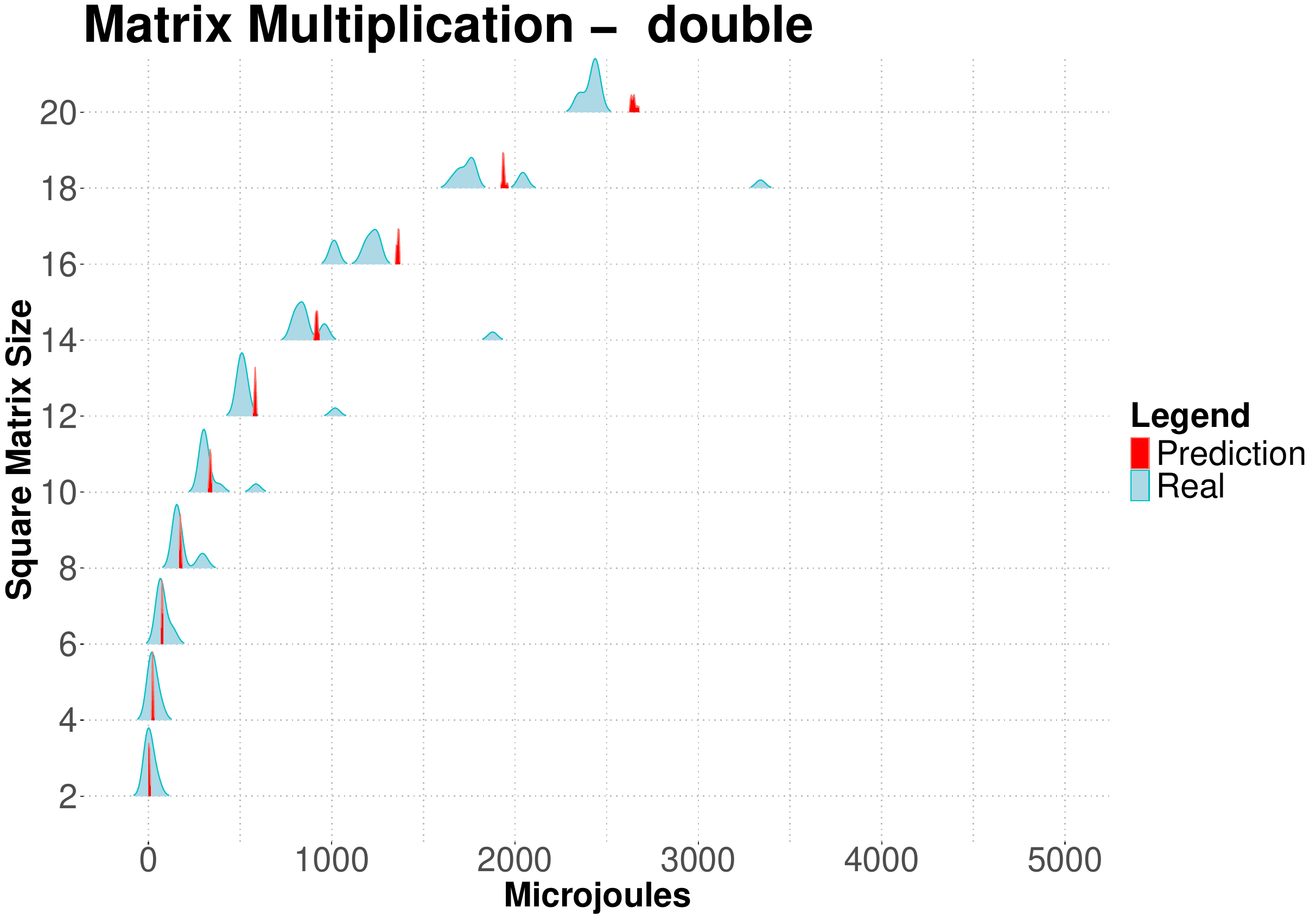} }
    \label{fig:predDouble}
    }%
    \subfloat[\centering Matrix Multiplication \code{float}]{{\includegraphics[width=.48\linewidth]{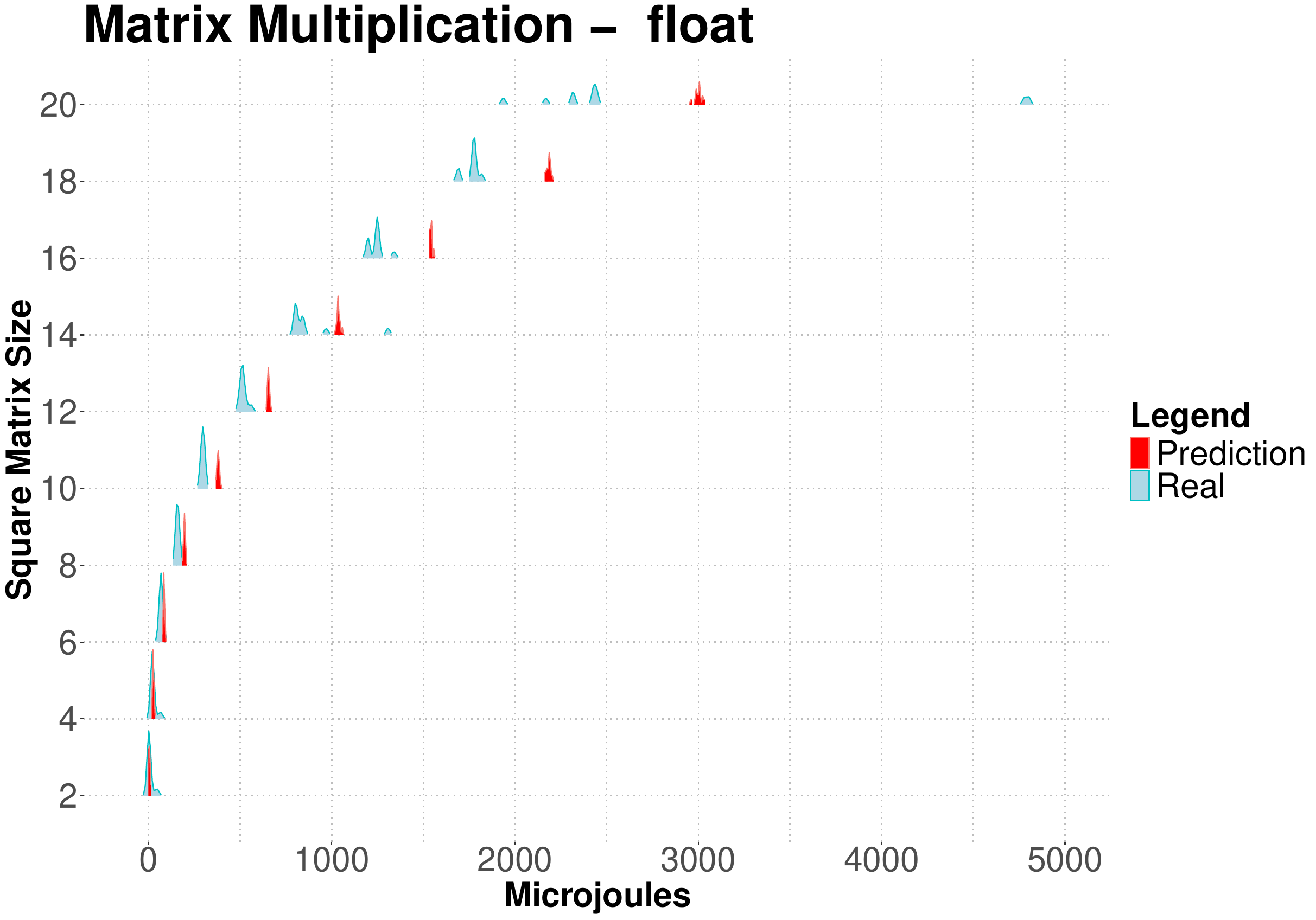} }
    \label{fig:predFloat}
    }%

    \caption{Energy predictions for matrix multiplication}
    \label{fig:pred_matrix}
\end{figure}

\begin{figure}
    \centering
    \includegraphics[width=.75\linewidth]{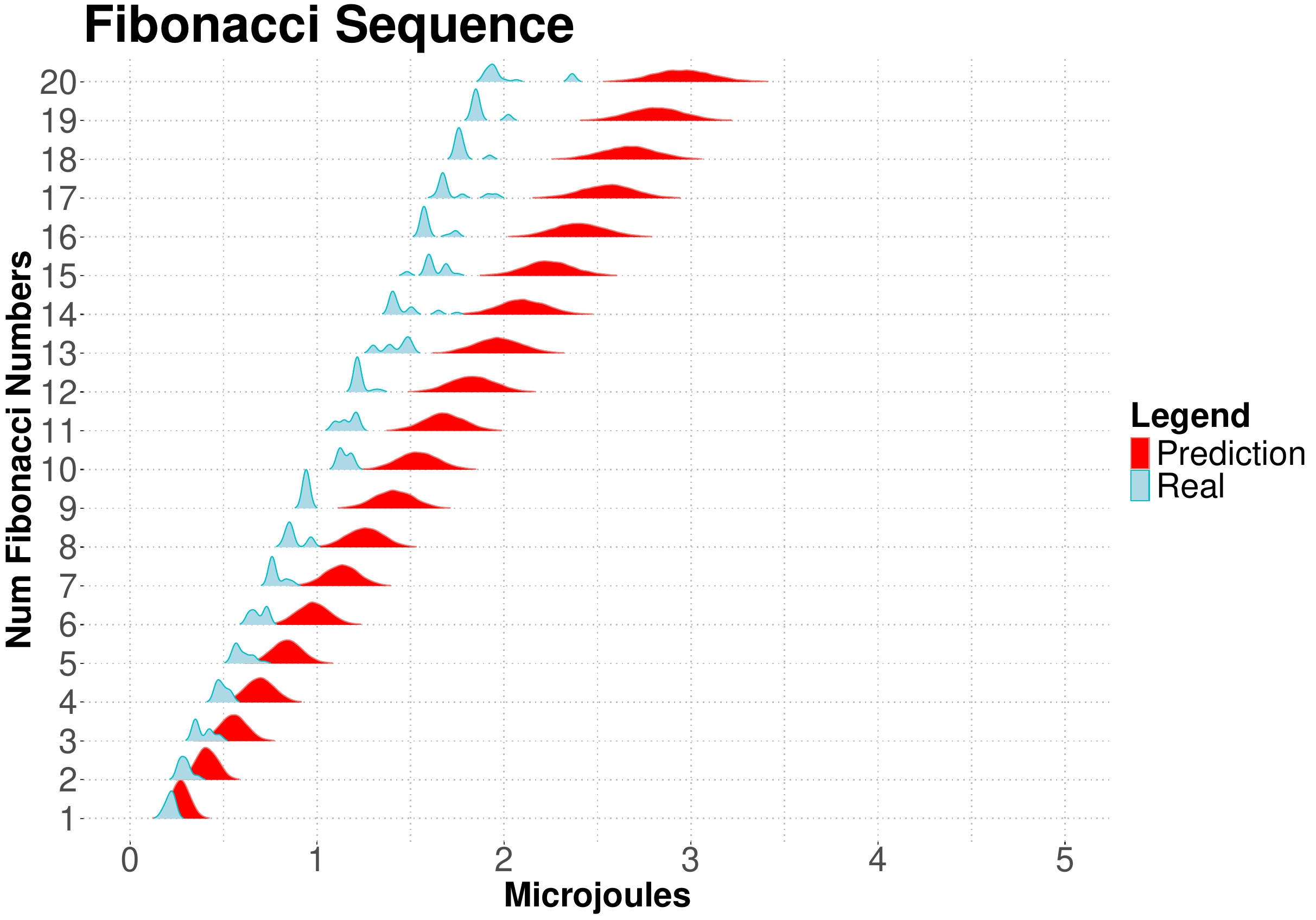}
    \caption{Fibonacci sequence predictions}
    \label{fig:pred_Fib}
\end{figure}

We measure the matrix multiplication for four matrices (see Fig.~\ref{fig:pred_matrix}), one for each data type, and the calculation of $N$ values of the Fibonacci sequence for the \code{long} data type (see Fig.~\ref{fig:pred_Fib}).
Our predictions are consistent with or slightly overestimate the energy consumption of the programs.
For the matrix multiplication the prediction's overestimation increases as the square matrix size increases.
However, the variation in real energy consumption increases with larger matrix sizes, possibly because larger matrix sizes have a larger impact on the memory state, hence leading to a variation increase.
For the Fibonacci sequence, all predictions slightly overestimates the real energy consumption.

\section{Discussion}
\label{sec:discussion}
The results show several trends that we discuss below.

\textbf{Data sizes and types}: Overall, energy consumption increases as the number and size of memory accesses increases:
loading from Java's constant pool requires an indirect load, and we observe that the cost of this load is significantly higher for 64-bit values than for 32-bit values.
Accesses to local variables consume comparatively little energy, likely because they have a high probability of residing in the L1 data cache and because they can be accessed via direct load and store instruction.
This hypothesis is consistent with the negligible energy consumption for \code{const} values, i.e., immediate operands, which are encoded directly in the bytecode instructions.
For data types, we do not observe a similarly clear correspondence:
specifically, our model attributes significantly lower energy consumption to floating point operations on (64-bit) \code{double} values than on (32-bit) \code{float} values, which we consider counter-intuitive.
We do not currently have a clear explanation for this behavior of the model, though it matches the majority of our measurements, where e.g.\@ arithmetic on \code{double} consumes less energy than on \code{float} values.
Our model associates operations on objects (\code{ref}) with a substantially higher energy consumption, compared to primitive types, though the significance of this difference is limited to a small set of operations shared between primitive and object types, namely local variable load/store and equality comparison operations.

\textbf{Hardware Instances}: We observe a noticeable difference in consumption between different hardware instances of the same device model, and accounting for this difference creates a more generalizable final energy consumption model.

\textbf{Operations}:
\emph{Arithmetic operations and bit-operations}, in this category, two operations have a lower energy consumption than the other operations in this category: negation and increase (see Fig.~\ref{fig:ModelRes_op_arith}). In contrast to all other bit and arithmetic operations, which pop two values from the operand stack and push the result of their computation to the stack, the negation and increase operation do not~\cite[Chap. 6]{JVMSpec}. The negation only pops one value, whereas the increase operation does not push or pop any value to the operand stack but instead updates a value in-place.
We hypothesize that energy consumption is directly dependent on the number of operand stack alterations.

Most \emph{Type conversion operations} consume between $0 - 0.1 \mu$J.
Six operations fall outside of this range, including all four conversions from floating point types to integer types.
These operations require conditional branches inside the bytecode interpreter to correctly handle the conversion to maximal and minimal \code{int} and \code{long} values as required by the Java language specification.
The other two conversions that stand out with a higher energy cost are \code{long} to \code{int} and \code{int} to \code{char}.
Despite manual inspection of the OpenJDK bytecode interpreter we are unable to identify a clear reason for this difference.

The \emph{Allocations, Method, Object, and Array operations} have significant differences, especially in allocation/declaration. Dynamic allocation operations consume the most energy in the study, whereas variable declarations/assignments consume much less.
Both dynamic allocation operations must allocate memory from the heap, which the JVM must initialize to its default values~\cite[Chap. 6.5]{JVMSpec}, i.e., set to zero.
By comparison, local variables are assigned an index in the local activation record at compile time~\cite[Chap. 2.6.1]{JVMSpec}; Java's Definite Assignment requirement ensures that local variable storage does not need to be pre-initialized.
These differences, together with the need for setting up object headers on the heap, may explain the corresponding differences in energy consumption.
For method calls, instance method invocations (indirect calls, due to dynamic dispatch) consume more energy than static method invocations (direct calls), as expected.
All field accesses consume similar amounts of energy, except for instance field load operations, which take substantially less energy than even static field load operations.
This difference may be due to better cache locality, either as an inherent property of dynamic field loads or as an artifact of our benchmarks.

For \emph{Control transfer operations}, all conditional branches (if-statement variations) cluster around 0.04 $\mu$J.
Since most typical applications for our methodology would not be able to predict whether branches will be taken or not, we similarly opted against distinguishing between branches that are taken vs.\@ branches that are not taken.

Our model constructs reliable estimations for the vast majority of the modelled Bytecode patterns. Whilst a few factors, such as the greater energy consumption of \code{int} than \code{double}, is counter-intuitive, most follow intuition and present accurate estimations. However, as George Box points out with his famous quote, ``\textit{All models are wrong, but some are useful}'', all models are built upon assumptions, have deficiencies, and do not entail that the underlying phenomenon is as simple as the model. Our model does not consider some factors in the runtime environment, such as temporal and spatial locality of memory accesses and stack size, or hardware considerations, such as cache architecture or CPU optimizations. All of this yields predictions that considerably simplifies the true behaviour of the Bytecode patterns' energy consumption for the given execution platform. Even though these factors obstruct the creation of perfect estimates, our probabilistic methodology provides an explainable statistical model that one can use for further analysis or as an energy model in various energy analysis tools.

\section{Related work}
\label{sec:relatedWorks}
There are two primary methods of obtaining measurement for an energy profiler: hardware- and software-based sampling. Hardware sampling uses external devices to measure the energy consumption of an application. Typical measurement devices are power meters~\cite{vlensApplications,APIEnergy,JVMPowerPerf,energybugs}, multimeters~\cite{distributedEnergy,powerscope}, and data acquisition systems (DAQ)~\cite{ANEPROF,oldJVMarticle,javaThreadsEnergy}. However, novel academic measuring systems, such as LEAP~\cite{LEAP}, GreenMiner~\cite{greenMiner}, and more~\cite{PowerPack,ecalc,vlens,elens,energySyscalls,HPCenergy} exist.

Software-based sampling utilizes data from the execution platform, such as hardware logs~\cite{petra} and performance counters~\cite{JVMcompilerFlags,HPCenergy}. Hence, this sampling method has access to various system states. While lacking the precision of hardware measurement, this method can be more accessible as it uses data directly from the device under test. One such technique is Intel's running average power limit (RAPL), an interface that reports the energy consumption of several system-on-chip power domains~\cite{intelRAPL}. Since our Raspberry Pi 5 uses an ARM Cortex-A76, this alternative was not available. Hence, we use external measurement equipment. 
Several parameters are important when defining an energy consumption model. The compiler and Java virtual machine (JVM) are essential when considering Java systems. Depending on which JVM one uses~\cite{JVMEnergy} and its execution options~\cite{JVMcompilerFlags}, the energy consumption varies, and the same is true for the compilation~\cite{compilerOptimisation}. Furthermore, one can alter the JVM through run-time options, where the garbage collector and JIT are central parts. Disabling JIT increases energy consumption~\cite{oldJVMarticle,JVMcompilerFlags,javaMemoryEnergy}, while using the default Garbage First Collector~\cite {GarbageCollectorG1} reduces energy consumption~\cite{JVMcompilerFlags}.

The measurements from the above mentioned methods, become input parameters for a statistical model. Previous work typically uses a summative model for a system's energy consumption. Some use a single factor in the model, such as processes or method execution~\cite{javaThreadsEnergy,petra,powerscope}, while others~\cite{ecalc,distributedEnergy,APIEnergy,vlens,HPCenergy} use multiple, either to increase accuracy or describe relationships between the factors under investigation.

\section{Conclusion}
\label{sec:conclusion}

This study presents a novel methodology for assessing statically typed JVM-based programming languages' worst-case energy consumption (WCEC). Achieved by using Java Bytecode patterns, the translation of a programming language's source code statement to its corresponding Java Bytecode representation. The methodology uses four factors, three of which can be inferred statically from the code: data size, data type, and operation type, and, since energy consumption depends on the underlying hardware, the last factor is the energy contribution of the hardware platform. These categories are modelled with Bayesian statistics, allowing for predicting unseen programs or functions with an energy distribution and explaining energy consumption differences attributed to these factors. 

We implement our methodology for Java to assess its accuracy and viability. The results show that the operation type affects the final energy consumption the most, followed by the data type, data size, and the device instance. Larger (64-bit) data sizes pushed to the operand stack consume more energy than smaller sizes. The long and float data types consume more energy than int and double. Finally, there is a noticeable difference in energy consumption when executing on different instances of the same hardware platform. 

With the Bayesian model, we show that our model can closely predict the WCEC of matrix multiplication and calculating $N$ values of the Fibonacci sequence, further strengthening the methodology's viability. Future work can expand the model to include the impact of different CPU and cache architectures, which likely significantly impact energy consumption. 
As our work defines a methodology for an energy model, not an automated tool, future work can use the energy model as a basis for program analysis or energy verification tools.


\printbibliography[title={References},heading=subbibliography]

\newpage
\section*{Appendix}
\setcounter{section}{0}
\renewcommand*{\thesection}{\Alph{section}.}
\section{Energy Model's Point Estimates}
\label{sec:appendix:a}
{\scriptsize\tabcolsep=2pt
    \begin{longtable}{ll|lllll}
    \caption{Point estimates in Joules}\\
    \toprule
    \textbf{Category} & \textbf{Description} & \textbf{Mean} & \textbf{MCSE}  & \textbf{Std.dev} & \textbf{ESS} &\textbf{$\hat{R}$} \\
    \midrule
    \makecell{\textbf{Model's Std. dev.}\\(see Eq.~\ref{eq:modelFormula_a})} & $\sigma$    & 1.356665e-08 & 5.704377e-13 & 7.316405e-11 &    16450.497  &  0.9998925 \\
    \hline
    \multirow{4}{1.5cm}{\textbf{Data size}} & 32-bit & 6.777943e-09 & 4.226145e-12 & 3.882128e-10 &     8438.222  &  1.0006578 \\
    & 64-bit & 1.072327e-08 & 4.006433e-12 & 3.860974e-10 &     9287.052  &  1.0000462 \\
    & Constant  & 5.287443e-11 & 3.124615e-13 & 3.718432e-11 &    14162.065  &  0.9996307 \\
    & Load & 2.972453e-10 & 1.979835e-12 & 2.223760e-10 &    12615.887  &  0.9999100 \\
    & Reference   & 9.362167e-09 & 1.712994e-10 & 1.000466e-08 &     3411.090  &  1.0015461 \\
    \hline
    \multirow{50}{1.5cm}{\textbf{Operation}}
    & Addition &  2.954780e-08 & 4.513539e-12 & 4.479418e-10 &     9849.377  &  1.0002523 \\
    & Division & 2.828019e-08 & 4.411272e-12 & 4.450900e-10 &    10180.474  &  1.0001077 \\
    & Increase & 4.841576e-09 & 1.420322e-11 & 1.404849e-09 &     9783.312  &  1.0000205 \\
    & Multiplication & 3.174836e-08 & 4.572187e-12 & 4.485468e-10 &     9624.266  &  1.0003379 \\
    & Negation & 4.156202e-09 & 7.048654e-12 & 7.634754e-10 &    11732.153  &  1.0001635 \\
    & Subtraction & 2.613292e-08 & 4.444976e-12 & 4.395197e-10 &     9777.276  &  1.0003361 \\
    & Modulo & 3.355088e-08 & 7.852670e-12 & 7.609978e-10 &     9391.437  &  1.0005535 \\
    \cmidrule(l){2-7}
    & Bit and &  2.752216e-08 & 1.014659e-11 & 1.036892e-09 &    10443.040  &  1.0000481 \\
    & Bit complement &  2.708684e-08 & 1.033120e-11 & 1.028558e-09 &     9911.869  &  0.9996481 \\
    & Bit or &  1.981219e-08 & 9.408900e-12 & 1.028813e-09 &    11956.263  &  0.9999245 \\
    & Bit xor &  2.298810e-08 & 9.839752e-12 & 1.031443e-09 &    10988.084  &  1.0002514 \\
    & Left bitshift & 1.684052e-08 & 8.354586e-12 & 1.023943e-09 &    15021.092  &  0.9999660 \\
    & Right bitshift & 1.694766e-08 & 8.642389e-12 & 1.034805e-09 &    14336.701  &  0.9995472 \\
    & Logical right bitshift & 1.667414e-08 & 8.565198e-12 & 1.016775e-09 &    14092.064  &  0.9994074 \\
    \cmidrule(l){2-7}
    &d2f &  3.939552e-09 & 1.474987e-11 & 1.383364e-09 &     8796.230  &  1.0003774 \\
    &d2i & 5.990642e-08 & 1.428061e-11 & 1.413998e-09 &     9804.012  &  0.9998444 \\
    &d2l & 6.144221e-08 & 1.404236e-11 & 1.399201e-09 &     9928.414  &  0.9997871 \\
    \cmidrule(l){2-7}
    &f2d & 5.701561e-10 & 3.872877e-12 & 4.874727e-10 &    15842.848  &  0.9994983 \\
    &f2i & 5.418630e-08 & 1.474614e-11 & 1.439671e-09 &     9531.678  &  0.9997629 \\
    &f2l & 5.397823e-08 & 1.431926e-11 & 1.405102e-09 &     9628.859  &  0.9997359 \\
    \cmidrule(l){2-7}
    &i2b & 5.231769e-09 & 1.367685e-11 & 1.396401e-09 &    10424.329  &  1.0001241 \\
    &i2c & 2.147922e-08 & 1.039942e-11 & 1.413092e-09 &    18463.873  &  0.9997138 \\
    &i2d & 3.358516e-09 & 1.434744e-11 & 1.361303e-09 &     9002.445  &  1.0001629 \\
    &i2f & 4.340072e-09 & 1.606212e-11 & 1.407451e-09 &     7678.230  &  1.0006056 \\
    &i2l & 8.443708e-09 & 1.121589e-11 & 1.380526e-09 &    15150.309  &  0.9998848 \\
    &i2s & 4.335038e-09 & 1.782580e-11 & 1.416223e-09 &     6311.981  &  1.0012059 \\
    \cmidrule(l){2-7}
    & l2d & 4.551390e-10 & 3.324453e-12 & 4.006268e-10 &    14522.443  &  0.9998190 \\
    & l2f & 5.300106e-10 & 3.946982e-12 & 4.670783e-10 &    14003.899  &  0.9997515 \\
    & l2i & 1.681959e-08 & 9.937823e-12 & 1.436558e-09 &    20896.041  &  0.9995970 \\
    \cmidrule(l){2-7}
    & Array length &  4.630274e-09 & 1.411479e-11 & 1.417610e-09 &    10087.061  &  1.0000091 \\
    & Array load &  9.386346e-09 & 5.694052e-12 & 6.984052e-10 &    15044.302  &  0.9995347 \\
    & Array store &  1.973851e-08 & 6.844493e-12 & 6.978155e-10 &    10394.382  &  0.9998606 \\
    & Array allocation & 4.166971e-07 & 7.258579e-12 & 6.855984e-10 &     8921.469  &  0.9999811 \\
    \cmidrule(l){2-7}
    & Object allocation & 4.251660e-07 & 1.419500e-11 & 1.468750e-09 &    10705.942  &  0.9995034 \\
    & Object get field & 1.163949e-08 & 6.323401e-12 & 7.694835e-10 &    14808.024  &  1.0000446 \\
    & Object get static field & 3.282841e-08 & 7.459323e-12 & 7.575560e-10 &    10314.082  &  1.0001719 \\
    & Object put field & 2.853163e-08 & 7.791865e-12 & 7.577485e-10 &     9457.303  &  1.0001011 \\
    & Object put static field & 3.130976e-08 & 7.695283e-12 & 7.748583e-10 &    10139.005  &  1.0001390 \\
    \cmidrule(l){2-7}
    & Static method call & 7.577298e-08 & 6.058233e-10 & 2.514026e-08 &     1722.057  &  1.0028088 \\
    & Non-static method call & 8.714123e-08 & 6.042395e-10 & 2.516451e-08 &     1734.437  &  1.0027943 \\
    & return statement & 1.474867e-07 & 7.908349e-12 & 7.698483e-10 &     9476.296  &  0.9996235 \\
    \cmidrule(l){2-7}
    & Switch-case consecutive & 1.423901e-08 & 1.055638e-11 & 1.406943e-09 &    17763.284  &  0.9996359 \\
    & Switch-case non-consecutive & 2.761723e-08 & 1.231643e-11 & 1.405279e-09 &    13018.341  &  0.9996373 \\
    \cmidrule(l){2-7}
    & if equal (references) & 3.195112e-10 & 2.355600e-12 & 2.825413e-10 &    14386.683  &  0.9994039 \\
    & if non-equal (references) & 1.001539e-10 & 6.151938e-13 & 7.821645e-11 &    16164.872  &  0.9998154 \\
    & if non-null (references) & 9.271004e-11 & 6.230011e-13 & 7.366605e-11 &    13981.609  &  1.0001024 \\
    & if null (references) & 3.951120e-10 & 2.983059e-12 & 3.552355e-10 &    14181.070  &  0.9996997 \\
    & if equal (numerical data types) & 3.766614e-08 & 7.855906e-12 & 7.598616e-10 &     9355.705  &  1.0000516 \\
    & if non-equal (numerical data types) & 3.561147e-08 & 7.595503e-12 & 7.493736e-10 &     9733.829  &  1.0003164 \\
    & if greater or equal (numerical data types) & 3.269893e-08 & 7.853031e-12 & 7.670695e-10 &     9541.019  &  0.9996652 \\
    & if greater (numerical data types) & 3.382851e-08 & 7.496544e-12 & 7.621013e-10 &    10334.828  &  0.9999051 \\
    & if less or equal (numerical data types) & 2.753095e-08 & 7.858564e-12 & 7.587121e-10 &     9321.110  &  0.9998173 \\
    & if less (numerical data types) & 2.918040e-08 & 7.672754e-12 & 7.606679e-10 &     9828.509  &  1.0002637 \\
    & if equal (int $\neq 0$) & 4.570354e-08 & 1.396288e-11 & 1.408371e-09 &    10173.817  &  0.9997026 \\
    & if non-equal (int $\neq 0$) & 3.623144e-08 & 1.379621e-11 & 1.417438e-09 &    10555.735  &  1.0004407 \\
    & if greater or equal (int $\neq 0$) & 5.294165e-08 & 1.401972e-11 & 1.413106e-09 &    10159.469  &  0.9996097 \\
    & if greater (int $\neq 0$) & 4.262262e-08 & 1.376264e-11 & 1.395319e-09 &    10278.833  &  1.0000008 \\
    & if less or equal (int $\neq 0$) & 3.619356e-08 & 1.379409e-11 & 1.404777e-09 &    10371.193  &  0.9998488 \\
    & if less (int $\neq 0$) & 3.510509e-08 & 1.370255e-11 & 1.421024e-09 &    10754.743  &  0.9997882 \\
    & else branch & 2.753481e-08 & 1.207184e-11 & 1.396459e-09 &    13381.644  &  0.9993639 \\
    \cmidrule(l){2-7}
    & Variable declaration & 7.106552e-09 & 3.652931e-12 & 4.426781e-10 &    14685.649  &  0.9996167 \\
    \hline
    \multirow{5}{1.5cm}{\textbf{Data type}} & \code{double} & 6.478463e-11 & 3.744616e-13 & 4.704906e-11 &    15786.552  &  0.9997915 \\
    & \code{float} & 5.437049e-09 & 3.012367e-12 & 3.150217e-10 &    10936.172  &  1.0000247 \\
    & \code{int} & 9.716775e-11 & 6.263412e-13 & 7.562467e-11 &    14578.237  &  0.9997455 \\
    & \code{long} & 6.986588e-09 & 3.033201e-12 & 3.111745e-10 &    10524.598  &  0.9998161 \\
    & Reference & 4.701721e-08 & 6.152794e-12 & 6.163997e-10 &    10036.451  &  0.9999792 \\
    \hline
    \multirow{2}{1.5cm}{\textbf{Device}}
    & Device 1 & 6.739114e-09 & 2.147099e-12 & 2.061548e-10 &     9218.980  &  0.9998192 \\
    & Device 2 & 5.074944e-11 & 3.040083e-13 & 3.596514e-11 &    13995.636  &  0.9994804 \\
    \bottomrule %
    \end{longtable}
}

\end{document}